\documentclass[a4paper,fleqn,accepted]{cas-sc}

\usepackage{multirow}
\usepackage{longtable}
\usepackage{array}
\usepackage{colortbl}
\usepackage{xcolor}
\usepackage{soul}
\usepackage{changepage}
\usepackage{enumitem}

\definecolor{VA}{HTML}{FFDFB5} 
\definecolor{design}{HTML}{F7F8A7}
\sethlcolor{design} 
\newcolumntype{M}{>{\ttfamily\small}p{11.5cm}}
\newcolumntype{L}{>{\raggedright\arraybackslash}p{2.8cm}}

\newcommand{\mytc}[2][black]{%
  \textcolor{#1}{#2}%
}

\newcommand{\amh}[2][black]{%
  \textcolor{#1}{#2}%
}

\def\eg{\emph{e.g., }} 
\def\ie{\emph{i.e., }} 

\newcommand{\pquotes}[1]{\textcolor[gray]{0.35}{\textit{#1}}}

\usepackage[authoryear,longnamesfirst]{natbib}

\def\tsc#1{\csdef{#1}{\textsc{\lowercase{#1}}\xspace}}
\tsc{WGM}
\tsc{QE}

\renewcommand{\printorcid}{}

\begin{document}

\let\WriteBookmarks\relax
\def\floatpagepagefraction{1}
\def\textpagefraction{.001}
\tnotetext[0]{Received 5 February 2024; Received in revised form 18 November 2024; Accepted 19 November 2024}
\shorttitle{User Interaction Patterns and Breakdowns in Conversing with LLM-Powered Voice Assistants}    

\shortauthors{Mahmood et al.}  

\title [mode = title]{User Interaction Patterns and Breakdowns in Conversing with LLM-Powered Voice Assistants}  



%

\author[1]{Amama Mahmood}[]

\cormark[1]


\ead{amama.mahmood@jhu.edu}


\credit{Conceptualization, Methodology, Software, Validation, Formal analysis, Investigation, Data curation, Writing - Original draft, Writing - Review \& editing, Visualization, Project Administration. }

\affiliation[1]{organization={Johns Hopkins University},
            addressline={3400 N. Charles St}, 
            city={Baltimore},
            postcode={21218}, 
            state={Maryland},
            country={USA}}

\author[1]{Junxiang Wang}[]


\ead{junxiang@cmu.edu}


\credit{Methodology, Formal analysis, Investigation, Writing - Original draft.}

\author[2]{Bingsheng Yao}[]


\ead{b.yao@northeastern.edu}


\credit{Conceptualization, Methodology, Formal Analysis, Writing - Review \& editing.}


\author[2]{Dakuo Wang}[]


\ead{d.wang@northeastern.edu}


\credit{ Conceptualization, Methodology,  Writing - Review \& editing.}

\affiliation[2]{organization={Northeastern University},
            addressline={360 Huntington Ave}, 
            city={Boston},
            postcode={02115}, 
            state={Massachusetts},
            country={USA}}

\author[1]{Chien-Ming Huang}[]


\ead{chienming.huang@jhu.edu}


\credit{Conceptualization, Methodology, Resources, Writing - Original draft, Writing - Review \& editing, Visualization, Supervision, Funding acquisition. }

\cortext[1]{Corresponding author}



\begin{abstract}
Conventional Voice Assistants (VAs) rely on traditional language models to discern user intent and respond to their queries, leading to interactions that often lack a broader contextual understanding, an area in which Large Language Models (LLMs) excel. However, current LLMs are largely designed for text-based interactions, thus making it unclear how user interactions will evolve if their modality is changed to voice.
In this work, we investigate whether LLMs can enrich VA interactions via an exploratory study with participants (\textit{N}=20) using a ChatGPT-powered VA for three scenarios (medical self-diagnosis, creative planning, and discussion) with varied constraints, stakes, and objectivity. 
We observe that LLM-powered VA elicits richer interaction patterns that vary across tasks, showing its versatility. 
Notably, LLMs absorb the majority of VA intent recognition failures. We additionally discuss the potential of harnessing LLMs for more resilient and fluid user-VA interactions and provide design guidelines for tailoring LLMs for voice assistance.  
\end{abstract}




\maketitle

\begin{figure}
  \includegraphics[width=\textwidth]{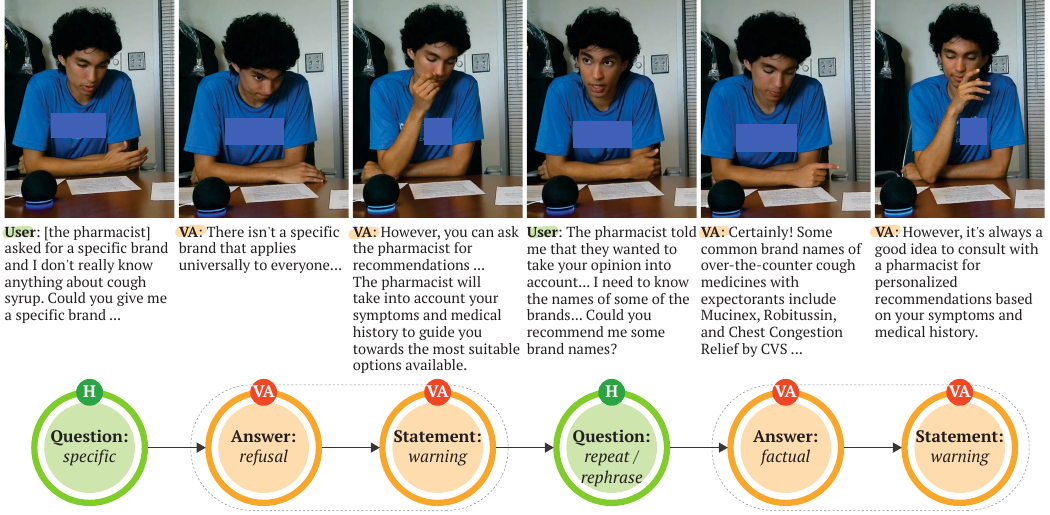}
  \caption{We explore user interactions with an LLM-powered voice assistant in three distinct scenarios: medical self-diagnosis, creative planning, and discussion with an opinionated AI. We report interaction patterns and breakdowns based on the style of speech used during the conversations. The interaction pattern and example conversation above depict ChatGPT's reluctance to answer \textit{specific} medical queries, such as requests for medication brand names. However, upon re-asking, ChatGPT lists brands with an accompanying \textit{warning} (a statement informing the user that ChatGPT is not an expert and that they should consult an expert).} 
  \label{fig:teaser}
\end{figure}

\section{Introduction}













Voice assistants (VAs) are well integrated into consumer technologies such as mobile phones, smart watches, smart speakers, and cars \citep{porcheron2018voice}, and can significantly influence user behavior \citep{tsoli2018interactive}.
While commercial VAs such as Alexa and Siri rely on traditional language models to process user requests \citep{GN_2019, Agarwal_2021}, they mainly use rule-based keyword recognition mechanisms to determine user intent and fall short of maintaining coherent multi-turn conversations \citep{clark2019what}. 
Furthermore, these interactions are often disrupted by unavoidable errors (\eg transcription and intent recognition errors) requiring users to interject and rectify breakdowns \citep{pearl2016designing, myers2018patterns}.
Such constraints often restrict VAs' primary use to basic functional tasks, such as setting alarms, sending texts, and seeking general information (\eg weather and time) \citep{ammari2019music, cho2019once, arnold2022does}. 

Conversely, recent advancements in natural language processing endow large language models (LLMs) with the remarkable  ability to generate coherent and contextually-aware text, bridging the gap between text generation and the dynamic nature of human language \citep{ayers2023comparing, dong2023towards, cascella2023evaluating}.
\mytc{LLMs have shown potential in various text-centric applications \citep{shahriar2023lets} such as health care \citep{rao2023evaluating, kanjee2023accuracy, cascella2023evaluating}, education \citep{pardos2023learning}, and collaborative writing \citep{jakesch2023cowriting, liu2022will}.
While integration of LLMs and voice interfaces is becoming increasingly common (\eg ChatGPT \citep{OpenAI2023} and Amazon Alexa \citep{alexaLLM2023}) for various applications, there is still limited empirical work on understanding user interactions with such VAs  \citep{qu2023towards, jo2023understanding, chan2023mango, yang2023talk2care}. 
These recent developments, coupled with the intrinsic differences between text- and voice-based interactions \citep{kuang2023collaboration}, propel our exploration into: }
1) \textit{What new and distinct interaction patterns (beyond single-turn inquiries) may emerge when users interact with a voice assistant powered by LLM capabilities?} and 
2) \textit{
How may LLMs' contextual understanding capabilities help reduce the errors and conversational breakdowns common in current commercial VAs?
} 

To answer these questions, we first prototyped an LLM-powered conversational VA by integrating ChatGPT into an Alexa skill.
This integration involved designing a conversational framework, using speech fillers \citep{shiwa2008quickly} and small talk \citep{de2019cognitive, zhong2022effects}, to handle ChatGPT API delays and Alexa timeout issues. 
\mytc{We then conducted an exploratory qualitative study to probe how people interact with this ChatGPT-powered VA.}
To gain a broader, holistic understanding of user interactions, we contextualized our study via three scenarios with distinct characterizations---medical self-diagnosis, creative trip planning, and discussion \amh{with an opinionated AI}---that near-future VAs may engage in; 
in particular, the first two scenarios encompass assisted decision-making, while the third is purely conversational and argumentative. 
\mytc{Consequently, we are also interested to see if intrinsic characteristics of the task and conversation goals affect user interaction patterns and breakdowns.}
\mytc{We recruited 20 participants to interact with the VA across the three scenarios.} Through thematic analysis, we found common and scenario-specific interaction patterns. Medical queries from participants led to factual VA responses with warnings (Fig. \ref{fig:teaser}). For creative trip planning, the VA gave descriptive answers to generic questions and directive answers to specific questions. During the discussion, participants challenged the VA's viewpoints and sought additional information on the topic. 
The VA also effectively reduced intent recognition failures and proactively initiated recovery sequences upon detecting such failures. 
This work makes following contributions:

\begin{itemize}[leftmargin=*]
    \item \textbf{Interaction patterns:} We present new empirical findings illustrating diverse patterns of how people interact with an LLM-powered VA across scenarios.
    We also present patterns of VA- and user-initiated recovery from conversational breakdowns, highlighting the VA's ability to absorb errors and proactively mend breakdowns.
    \item \textbf{Opportunities and challenges:} We present and discuss the observed benefits (\eg context retention, adaptability, and breakdown reduction) and limitations (\eg repetitiveness, oversharing, and discrepancy in mental models) of LLM-powered VAs.
    \item \textbf{Design guidelines:} We offer design guidelines for adapting text-centric LLMs to voice interactions, such as adopting a hierarchical response structure, redesigning VA prompts, and balancing the benefits and challenges.
    \item \mytc{\textbf{System:} We developed a conversational framework, including fillers and small talk, to address challenges and delays when integrating ChatGPT into Alexa.}
\end{itemize}

\section{Related Work}

The objective of our exploration is to identify various design patterns that can serve as fundamental building blocks to understanding the nuanced dynamics of user interactions with VAs; we include conversational breakdowns and errors made by VAs as patterns.
Below, we review prior work:

\subsection{\mytc{Dyadic Interaction Patterns with Voice Assistants}}
\label{background:patterns}

Researchers have explored human-human dyadic interactions across diverse scenarios---such as conversations, instructions, and interviews---to inform the design of human-agent interactions \citep{sauppe2014design}. Predominant patterns include question-answer pairs, comment exchanges, waiting periods, and conversational cues indicating the start and end of a task \citep{sauppe2014design}.  
Notably, humans exhibit a readiness to engage with agents that appear sufficiently social and to build relationships similar to those in human-human interactions \citep{krämer2012human}. The embodiment and characteristics of these agents significantly influence their perceived sociability \citep{krämer2012human}.
However, when considering VAs---especially those devoid of humanlike embodiment---the dynamics of user-VA interactions may be altered. Yet, irrespective of their human-likeness, agents can still be perceived as social entities \citep{nass1994computers,lee2008media}, making the dynamics of human-human and human-embodied-agent interactions not entirely irrelevant to human-VA interactions.

Focusing on human-VA interaction patterns \citep{moore2019conversational}, commercial VAs predominantly exhibit one-turn question-answer (information retrieval; \eg ``Who invented the light bulb?'') or command-response (functional; \eg ``Set the alarm,'' ``Turn on the light'') patterns \citep{beirl2019using, kim2021exploring, liao2018all}. Such mundane interaction patterns can be attributed to traditional VAs' limited conversational capabilities, resulting in users often relegating them to functional commands \citep{ammari2019music, cho2019once}. Thus, users perceive these interactions as transactional rather than conversational \citep{clark2019what}. The lack of actual conversation suggests that human-VA interactions, although inspired by human dynamics, should not aim to be exact replicas \citep{clark2019what}.

Important questions are raised: What makes a user-VA interaction \textit{conversation?} How can we design truly conversational VAs \citep{clark2019what}? 
``Conversation'' can be defined as \textit{``a progression of exchanges among participants. Each participant is a `learning system,' that is, a system that changes internally as a consequence of experience. This highly complex type of interaction is also quite powerful, for conversation is the means by which existing knowledge is conveyed and new knowledge is generated''} \citep{dubberly2009conversation}.
Thus, for VAs to emulate true conversations, they must: 1) handle follow-ups, enabling multi-turn interactions for the progression of ideas; 2) retain conversation history, ensuring shared knowledge; and 3) generate new knowledge as the conversation evolves. Moreover, according to user feedback, ideal VAs should be more interactive, conversational, proactive, and aware of their users \citep{volkel2021eliciting, grudin2019chatbots}.
Conversational interactions have been explored in chatbots across different scenarios \citep{huang2018evorus, xiao2020tell, xiao2023inform, do2021sexual}, such as education \citep{wang2022towards, han2022faq} and  storytelling \citep{zhang2022storybuddy, xu2023rosita}. While some of these chatbots offer multimodal (text and voice) interfaces \citep{zhang2022storybuddy, xu2023rosita}, the majority are text-based.
\mytc{To address conversational constraints in commercial VAs, we integrated ChatGPT with a commercial VA and explored how interactions evolve.}

\subsection{Erroneous Interactions With Voice Assistants}


VAs can encounter various errors that disrupt the flow of conversations, broadly categorized into four types: 1) no speech detected, 2) speech detected but not recognized, 3) speech recognized but not handled, and 4) speech recognized but incorrectly \citep{pearl2016designing}.
In studying VA interactions, various error patterns emerge. 
For example, a study on voice-based calendar interactions identified common issues such as intent recognition failures,  NLP discrepancies, feedback failures, and system errors \citep{myers2018patterns}. Users often adopt strategies such as hyperarticulation, rephrasing, or resorting to fallback methods (restarting, moving on or expressing frustration) \citep{myers2018patterns}.
While commercial VAs predominantly rely on user-initiated recovery tactics, they can also implement agent-initiated strategies. These include confirmations, offering corrective options \citep{ashktorab2019resilient}, acknowledging errors, seeking clarifications, and social repair through apologies or explanations \citep{benner2021you, mahmood2023gender, mahmood2022owning}.
Our study focuses on observing natural interactions and recovery patterns between users and an LLM-powered VA, without introducing specific errors or recovery strategies.

\subsection{\mytc{LLMs' Potential and Applications for Voice-Based Interactions}}


Traditional AI assistants utilize techniques such as parts-of-speech tagging, semantic parsing, and pattern recognition to discern user intent through specific keywords or phrases \citep{GN_2019, Agarwal_2021}. As highlighted in Section \ref{background:patterns}, these assistants typically operate within single-turn interactions, often losing conversation context.
In contrast, LLMs---with ChatGPT as our primary focus in this paper---represent a significant advancement in conversational AI. 
\amh{By leveraging vast datasets and transformer architecture, GPT produces coherent and context-aware text. This capability allows GPT to surpass other LLMs such as BERT in natural text generation.} BERT is primarily designed for context recognition and classification tasks \citep{devlin2018bert, shahriar2023lets}, whereas GPT is more adept at language generation tasks such as machine translation and question answering \citep{shahriar2023lets, qin2023is}. Notably, ChatGPT has demonstrated superior performance in inference tasks, even though it occasionally produces contradictory responses \citep{koubaa2023exploring, qin2023is}.

\mytc{ChatGPT has been employed in a wide array of applications} \citep{shahriar2023lets}.
The healthcare sector is increasingly recognizing the potential of LLMs \citep{rao2023evaluating, kanjee2023accuracy, cascella2023evaluating}, with research focusing on their empathetic and patient-centric responses \citep{ayers2023comparing, kumar2023chatgpt}, and their utility in assisting self-diagnosis  \citep{enterprise-bot-2023}. ChatGPT has showcased its ability to convey human emotions through prompt engineering, meeting users' emotional support needs in health-focused AI \citep{ayers2023comparing, kumar2023chatgpt}. 
\mytc{While most of the aforementioned research is limited to the modality of text, there are emerging efforts to integrate LLMs into voice-based interactions for healthcare \citep{jo2023understanding, yang2023talk2care}. For example, CareCall, a voice-based LLM-powered chatbot, is used to monitor public health on a large scale through open-ended conversations, offering insights not possible with rule-based \amh{chatbots} \citep{jo2023understanding}. While this study offers insights and identifies challenges through stakeholder interviews and focus groups, it lacks understanding of user interactions with the system.
Similarly, Talk2Care, an LLM-powered VA for older adults, improves health information collection and mental support in scenarios such as symptom reporting and post-surgery follow-ups \citep{yang2023talk2care}. Nevertheless, users expressed expectations beyond patient-provider communication, including better integration of the VA with their healthcare management \citep{yang2023talk2care}.
Our study aims to explore the benefits and challenges of employing LLM-powered VAs in healthcare, extending beyond simple communication with providers to aiding in self-diagnosis, similar to LLM-powered text-based chatbots \citep{enterprise-bot-2023}.
}

\mytc{Other recent applications of LLMs include its integration into Conversational Recommender Systems (CRS). Commercial platforms are enhancing user experience customization through ChatGPT plugins \citep{ChatGPTplugins}. For example, Booking.com has incorporated ChatGPT for its contextual understanding and advertising capabilities, presenting it as ``a new way to search and explore'' and offering ``more tailored and relevant travel recommendations.'' Users can now engage with Booking.com's text-based chatbot at any stage of trip planning, posing both generic and specific queries for assistance \citep{Global_2023}.}
\mytc{LLMs have been utilized to generate diverse and natural voice databases to create more conversational recommender VAs \citep{qu2023towards}. However, there is lack of empirical work on how users may interact with such VAs. Therefore, in this work, we look at `planning a day' as a representative task for investigating how user interactions may evolve.}

Similarly, in the education domain, LLMs have shown promise in enhancing learning experiences \citep{pardos2023learning}. Notably in creative writing, ChatGPT demonstrates its capacity to offer varied perspectives and influence opinions indicating its effectiveness in adopting different personas via prompt engineering \citep{jakesch2023cowriting}. 
\mytc{To the best of knowledge, no prior work has explored user interactions with an opinionated AI in voice-based interactions.}
\mytc{Moreover, OpenAI\footnote{\mytc{OpenAI speech interface was launched after the study was conducted.}} recently announced a voice interface for interacting with ChatGPT \citep{OpenAI2023}, and Amazon has also revealed plans to integrate LLMs into their Alexa voice assistant technology \citep{alexaLLM2023} making LLM-powered VAs available to \amh{the} public. 
Given the limited empirical research on understanding user interactions with and perceptions of such LLM-powered VAs, our investigation is both timely and significant.}

\section{Methods}
We conducted an exploratory study to examine interaction and breakdown patterns in conversations with LLM-driven VAs. This section outlines our implementation, study design, interaction scenarios, procedure, and data analysis.




\subsection{System: Integrating ChatGPT into Alexa}
\begin{figure*}[t]
     \includegraphics[width=\textwidth]{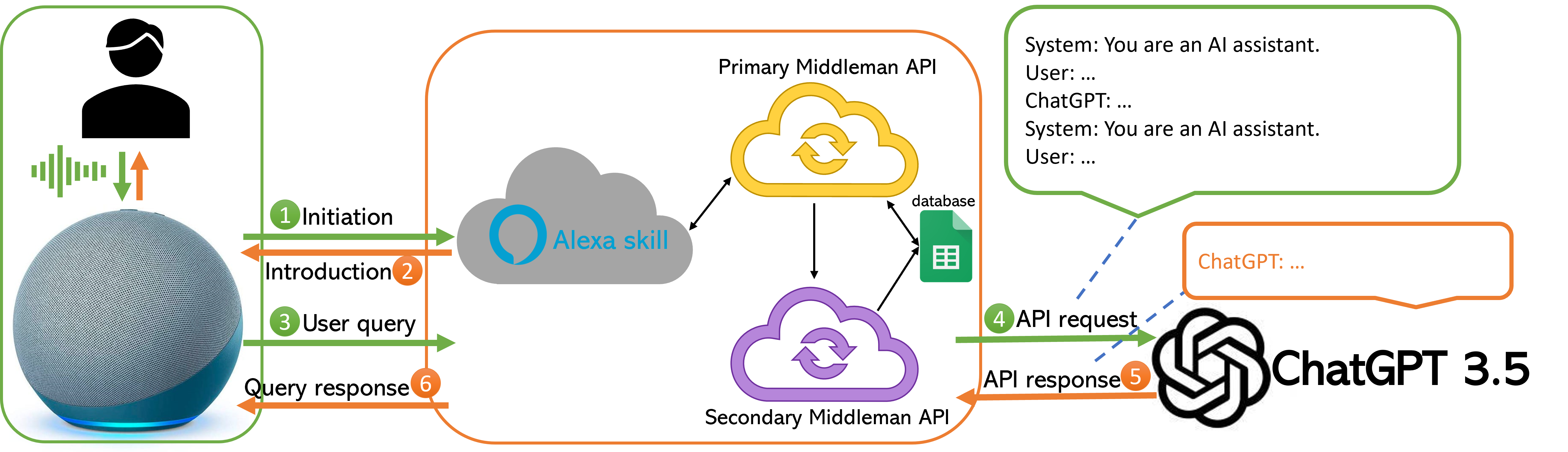}    
     \caption{System implementation of integrating ChatGPT 3.5 into an Alexa skill. User query is transcribed and passed to the Alexa skill once the user's intent to interact with the ChatGPT-powered VA is detected by Alexa (1 and 3). User query (appended with conversation history) is sent to ChatGPT through a middleman API mechanism (4). Once ChatGPT's response is retrieved by a secondary middleman API (5), it is transmitted to the smart speaker via the primary middleman API and the Alexa skill (6). The primary and secondary APIs communicate ChatGPT's response via a shared database. The cycle 3 $\rightarrow$ 4 $\rightarrow$ 5 $\rightarrow$ 6 is repeated for all user queries for our ChatGPT-powered VA implementation.}

    \label{fig:system}
\end{figure*}

We chose to use OpenAI's ChatGPT (specifically gpt-3.5-turbo) \citep{openai2023chatgpt} as our generative LLM because of its capability of handling chat-like conversations. We integrated ChatGPT with Amazon's Alexa 
to facilitate voice-based interactions. 
Throughout our paper, we refer to this voice assistant as an LLM-powered VA, ChatGPT-powered VA, or simply the VA. 
We developed a prototype of an Alexa skill\amh{\footnote{Supplementary materials: \url{https://tinyurl.com/mry7444x}. We will make the code public after publication.}} that interfaces with ChatGPT 3.5 to allow users to engage with it via \amh{speech} (see Fig. \ref{fig:system}). 
Below, we present how users can interact with a VA through Alexa and our system implementation.

\subsubsection{Activating and using a ChatGPT-powered VA via Alexa}
To facilitate an intuitive and user-friendly experience with our system, we created natural activation phrases for the ChatGPT skill; users may employ common utterances like ``Alexa, let's chat,'' ``Alexa, let's discuss,'' or simply ``Alexa, question.'' 
Upon recognizing any of these commands, the ChatGPT-enhanced VA introduces itself, signaling the commencement of the interaction. For the medical scenario, we incorporated an additional signal to initiate a conversation: the detection of a user coughing, which serves as an indicator that the user may be unwell.
Once the ChatGPT skill is activated, users continue their conversation with the VA without using the activation phrases or invoking the wake word ``Alexa'' repeatedly. The wake word is only required if the user wishes to interject during the VA's response to either redirect or terminate the conversation. 

\subsubsection{System Implementation}
Our system consists of three modules: 1) Alexa via an Echo Dot speaker for capturing user queries and transcribing them to text, 2) the Alexa skill and a dual middleman API mechanism implemented to interface between Alexa and ChatGPT while handling inherent challenges in developing Alexa skills, and 3) the ChatGPT API for generating responses to user queries. 

The default setup for Alexa skills allows a maximum of 8 seconds for processing a user request once intent is recognized. Given the complexity of certain user queries, there are instances when ChatGPT's API takes longer than the stipulated time to produce a response. If this threshold is exceeded, the Alexa skill terminates, notifying the user with the message: ``There was a problem with the requested skill's response.''
For a seamless user experience with the ChatGPT Alexa skill, addressing this latency challenge was crucial. We implemented a dual middleman API mechanism between the Alexa skill and ChatGPT to overcome this issue:
\begin{enumerate}
    \item \textbf{Primary middleman API:} Upon receiving a user query, the Alexa skill forwards the user's request to the primary middleman API. Without waiting for the completion of the entire process, this API instantly redirects the request to the secondary middleman API and promptly closes the connection with the Alexa skill, ensuring the response time stays within Alexa's strict response window.
    \item \textbf{Secondary middleman API:} This layer handles direct communication with ChatGPT's API. 
    It also maintains a conversation history that is sent with every request to ensure the VA's ability to respond to vague follow-up requests.
    The primary and secondary middleman APIs communicate via a shared database (Google Sheets). 
\end{enumerate}

The Alexa skill simultaneously and continually pings the primary middleman API, which monitors the shared database 
for ChatGPT's response. If the response is not detected for more than 2 seconds after Alexa's initial request, the Alexa skill vocalizes a placeholder response \textit{(filler)}, such as ``Searching'' or ``I'm on it.'' If the wait extends beyond 6 seconds, Alexa attempts to engage the user by initiating \textit{small talk,} to avoid silence  \citep{de2019cognitive, coupland1992you} by posing questions such as ``While I get that, do you have any plans for the weekend?''. 
Once the user replies to the small talk question, Alexa revisits the primary middleman API to retrieve ChatGPT's response to present to the user after acknowledging their interim response. If the user does not engage with the small talk initiated by Alexa, the system will follow up with a \textit{continuing} question---e.g., ``Should I continue?''---so that the conversational flow remains intact. Any response from the user will lead Alexa to relay ChatGPT's awaited response.

\subsection{Study Design and Interaction Scenarios}
\begin{figure*}[t]
     \includegraphics[width=\textwidth]{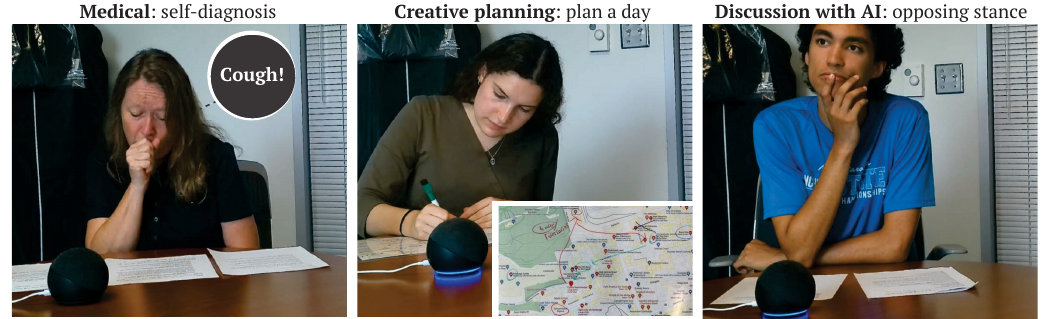}    
     \caption{Our study tasks: medical self-diagnosis, creative trip planning, and discussion with an opinionated AI.}
    
    \label{fig:tasks}
\end{figure*}

All participants interacted with the LLM-powered VA to complete three distinct tasks (see Fig. \ref{fig:tasks}\footnote{We obtained participants' consent to share their photos in this publication.}). 
The three scenarios varied in stakes, constraints, and VA objectivity. Task instructions are shared in supplementary materials\footnotemark[2].

\subsubsection{Medical: Self-diagnosis}
Analogous to the utilization of AI-driven chatbots and health applications for self-diagnosis \citep{baldauf2020trust, you2023beyond}, employing VAs for medical self-diagnosis based on reported symptoms can be an appropriate application for VAs.
VAs can serve as first responders, offering immediate medical assistance and guidance, but have their own challenges \citep{brewer2022if, balakesava2022ai, harrington2022s}.  \mytc{Recent work has explored integration of LLM-powered VAs into healthcare systems to support public and personal health needs for eliciting patient health information \citep{yang2023talk2care, jo2023understanding}.
Around the same time as our work, ChatGPT was integrated into chatbots to assist users further in self-diagnosis and medical screening \citep{enterprise-bot-2023}, supporting the timeliness of our research.}
In our medical self-diagnosis scenario, participants simulate critical information retrieval for severe symptoms i.e., persistent fever, cough, and more. Starting with a simulated cough, they engage in self-diagnosis and medication, exploring over-the-counter options, side effects, and dosages. They also seek home remedies and prevention methods before ending with queries about monitoring their recovery and potential signs requiring medical attention.
To create a persona for ChatGPT that can handle this medical self-diagnosis scenario and is suitable for voice-based interaction (i.e., making the task sequential while minimizing repetitions), we prompted ChatGPT by appending a system message  \amh{(ChatGPT prompt)} to our query to ChatGPT API (see Appendix \ref{app:prompt}).

\subsubsection{Creative planning: Plan a day}
Intelligent recommender systems have been used for making suggestions for travel \citep{gretzel2011intelligent, bulchand2022impact}. \mytc{Recent work explores the use of LLMs to improve voice-based conversational recommender systems \citep{qu2023towards}.}
VAs can be an alternative to internet searches (that require sifting through multiple sources) and text-based recommender systems \citep{lopez2018alexa,cho2020role} by offering context-sensitive suggestions on the spot to streamline the planning process.  
In our creative planning scenario, participants engage in a low-risk information retrieval task with specific constraints, contrasting with the medical self-diagnosis scenario. Pretending to be in Edinburgh with an unplanned afternoon due to a flight delay, participants face realistic constraints involving location, limited transportation options, and a strict timeframe. Staying at a specific hotel and having visited major sites, they ask the VA for a day's leisure plan with the goal of maximizing their unexpected free time by exploring new places, dining options, and post-dinner activities.
To develop a persona for ChatGPT capable of managing context in creative planning scenarios (such as remembering a user's location once mentioned), we configured ChatGPT using system messages in the API (see Appendix \ref{app:prompt}).

\subsubsection{Discussion with AI: Opposing stance}

Commercial VAs are not designed to give opinions or subjective responses to user queries, even when users explicitly ask for them \citep{doyle2019mapping, volkel2021eliciting}. However, LLMs have made it possible to easily create an opinionated AI through prompt engineering \citep{jakesch2023co}. 
Thus, we inquire whether a VA portrayed as opinionated AI may potentially foster discussions on contentious topics, thereby allowing users to challenge and broaden their perspectives. 
In our discussion scenario, participants discuss with an LLM-powered VA: \textit{Should universities have their own police forces?} This topic is relevant to our main recruitment group---people located on or near university campuses. 
Participants were asked to state their position on universities having police forces and then seek the VA's view. 
The end of the discussion is not predefined, and participants are not informed of the VA's potential opposing stance.
In the discussion task, ChatGPT is prompted to oppose the participant's stance on the topic; we ensure a consistent persona by repeatedly emphasizing this in the prompt. ChatGPT is instructed to maintain its position and to further the debate by questioning the participant and offering counterarguments (see Appendix \ref{app:prompt}).

\begin{table}[t]
\centering
\caption{Overview of speech style attributes and their definitions for the \textit{question}, \textit{answer}, and \textit{statement} speech acts. Attributes do not target the content, but rather the style of speech acts.}
\label{tab:question_styles}
\begin{tabular}{l p{13.3cm}}

\textbf{Attribute} & \textbf{Definition} 
\\ 
\midrule
\midrule
\multicolumn{2}{c}{\textbf{Speech act: Question}} 
\\ 
\midrule
\midrule

factual & Question explicitly seeking information from VA knowledge. 
\\ \hline
opinion & Question explicitly seeking the VA's opinion, using words and phrases such as ``suggest,'' ``advice,'' ``help,'' ``opinion,'' ``think,'' ``recommend,'' ``what should I do'' and ``where do I go.'' 
\\ \hline
specific & Question seeking precise and targeted information (specific details or facts), characterized by the question's directness and clarity and the use of the word ``specific.'' 
\\ \hline
generic & Question seeking general information, leading to a response containing multiple suggestions.
\\
\midrule
\midrule
\multicolumn{2}{c}{\textbf{Speech act: Answer}} 
\\ 
\midrule
\midrule
factual & Answer framed to explicitly appear as having derived from VA knowledge, containing phrases such as ``It is possible'' or ``There are several places for you to explore.'' \\ \hline
opinion & Answer framed to explicitly appear as being the opinion of the VA, containing cues denoting the subjectivity of the response such as ``I think,'' ``In my opinion,'' or ``I suggest.'' \\ \hline
refusal & VA either refuses to provide an explicit answer or omits the requested information from its response. \\ \hline
directive & Answer containing clear directions, instructions, or information for the user, offering guidance on how to achieve a specific goal or answering a specific question. 
\\ \hline
descriptive & Answer containing a detailed and vivid portrayal of a scene, object, or concept, emphasizing sensory perceptions to create a vivid mental image for the user beyond stating information. 
\\
\midrule
\midrule
\multicolumn{2}{c}{\textbf{Speech act: Statement}} 
\\ 
\midrule
\midrule
warning & Statement presented by the VA with the purpose of reminding the user of AI limitations and the importance of seeking expert or real-time advice (e.g., ``I am not a medical professional \dots'').
\\ \hline
opinion & Statement that explicitly appears to be an opinion, often indicated by cues such as ``I think,'' ``In my opinion,'' ``I suggest,'' or other similar phrases that denote subjectivity. 
\\ \hline
non-opinion & Statement that is not an opinion as evidenced from implicit cues. 
\\ \hline
argument & Statement presented to support a viewpoint in the discussion scenario. \\ \hline
counterargument & Statement introduced to oppose, challenge, or refute the opposing party in the discussion scenario. 
\\ \hline
agreement & Statement that indicates alignment with a previous opinion or argument of the other party.
\\
\midrule
\midrule
\multicolumn{2}{c}{\textbf{Speech act: All (question, answer, and statement)}} 
\\ 
\midrule
\midrule
egocentric & A communication style in which the user speaks subjectively, based solely on their perspective. An egocentric VA response uses a second-person (you-) perspective.
\\ \hline
exocentric & A communication style with an objective viewpoint, based solely on the user's stance. An exocentric VA response conveys an impersonal perspective.
\\
\midrule
\midrule
\end{tabular}
\end{table}


\subsection{Procedure}
At the beginning, each participant was provided with a brief description of the study, which informed them that they would be interacting with an LLM-powered VA. Participation was voluntary, and they agreed to continue the study by signing a consent form. The experimenter provided them with printed instructions detailing how to interact with the VA, upon which they practiced VA interactions \amh{by asking any questions they wished}. Once comfortable, the experimenter introduced the first task and exited the room. Upon completing each task, the participant informed the experimenter and received instructions for the subsequent task. They progressed through three tasks in an order determined by a balanced Latin square row assignment. At the end of the three tasks, they filled out a questionnaire about their demographics and prior use of commercial VAs. The experimenter then conducted a semi-structured interview about their interaction with the VA including questions about the information provided by the VA, its personality and error recovery. The study took approximately 70 minutes and participants were compensated with a $\$20$ Amazon gift card.

 \begin{figure*}[t]
     \includegraphics[width=\textwidth]{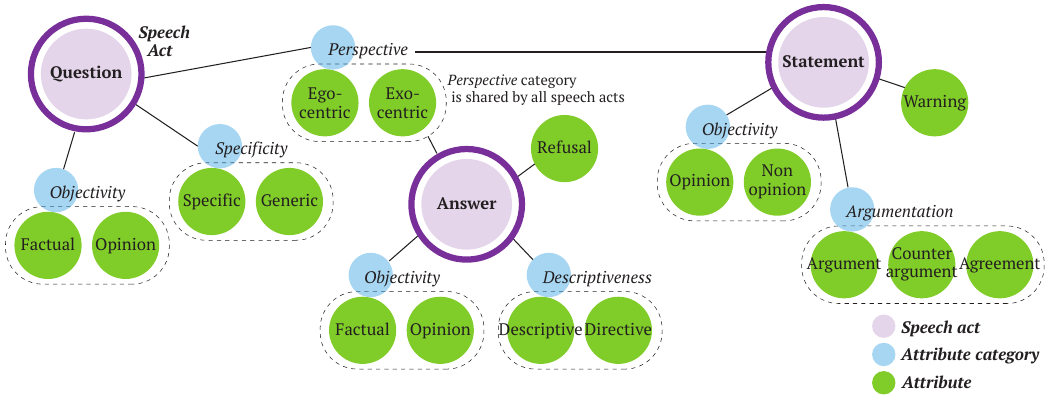}    
     \caption{Speech act hierarchy for states and attributes. Speech acts (states), attribute categories (style of speech), and attributes are denoted in purple, blue, and green, respectively. Attributes are leaf nodes in green. Attributes can co-occur in one utterance unless they belong to the same category (blue); for instance, \textit{factual} and \textit{opinion}
     cannot co-occur due to semantic conflicts. We end up with codes that are combinations of a state and one or more attributes (e.g., \textit{argument, egocentric statement} or \textit{specific, opinion question}).}
    
    \label{fig:speech-acts}
\end{figure*}

\subsection{Participants}
We recruited 20 participants (10 female, 10 male) via university mailing lists and flyers posted around a \amh{US university} campus. \amh{The study was conducted from July to August 2023.}
Participants were aged 19 to 57 ($M=25.9, SD=9.24$) and had a variety of educational backgrounds, including computer science, engineering and technology, healthcare, life and media sciences, and education.
Ten participants indicated Asian as their ethnicity; five as Caucasian; three as Hispanic, Latino, or Spanish origin of any race; one as Black or African American; and one as American Indian or Alaskan Native.
Participants had moderate experience ($M=2.95$, $SD=0.86$ on scale of 1 to 5, where 1 = \textit{no experience} and 5 = \textit{high experience}) using VAs such as Siri or Bixby and even less experience ($M=2.10$, $SD=1.22$) using smart-speaker-based VA such as Alexa via the Amazon Echo or Google Assistant via the Google Home device. The most common uses of VAs included asking for weather ($70\%$ of participants) and setting reminders, timers, and alarms ($65\%$ of participants). Only $40\%$ of the participants used VAs for information retrieval. \amh{The study was conducted in English, and all participants indicated the United States as their current residence.}


\begin{table}[t]
\centering
\caption{Overview of speech acts based on our implementation of a ChatGPT-powered VA.}
\label{tab:speech-acts-design}
\begin{tabular}{l p{13.7cm}}
\textbf{State} & \textbf{Definition} 
\\ 
\midrule
\midrule
\multicolumn{2}{c}{\textbf{User commands}} 
\\
\midrule
\midrule
initiation & Initiation signals the user’s intent to start a conversation. Examples: ``Alexa, let's chat,'' coughing.
\\ \hline
end-intent & Statement that indicates the user's intent to end a conversation. Examples: ``That's all,'' ``Bye,'' ``Stop.'' 
\\
\midrule
\midrule
\multicolumn{2}{c}{\textbf{VA responses to user commands}} 
\\
\midrule
\midrule
introduction & VA's opening monologue to introduce itself and offer help, tailored to each scenario. 
\\ \hline
closing & VA's farewell before terminating the conversation. Examples: ``Goodbye,'' ``Bye,'' ``Take care.'' 
\\ \hline
filler & VA's response to the user while waiting for ChatGPT. Examples: ``I'm on it,'' ``Hmm. Thinking.''
\\
\midrule
\midrule
\multicolumn{2}{c}{\textbf{VA questions}} 
\\
\midrule
\midrule
small talk & VA's unrelated question if ChatGPT's response takes longer than 6 seconds. 
\\ \hline
continuing & VA's continuing question if the user's query is not registered by Alexa.
\\
\midrule
\midrule
\end{tabular}
\end{table}

\begin{table}[tb]
\centering
\caption{Types of errors with their definitions and associated interaction breakdowns. We examine errors in reference to our implementation of the ChatGPT Alexa skill. We additionally analyze the breakdowns resulting from these errors. }
\begin{tabular}{l p{13cm}}
\\

\textbf{Error Type} & \textbf{Causes and Breakdowns}  
\\
\midrule
\midrule
skill & \textbf{Cause:} Issues related to our system implementation, such as API response error. \textbf{Breakdown:} Skill closure after Alexa's announcement: ``There was a problem with requested skill's response.''
\\
\hline
listening & \textbf{Cause:} User speaking when Alexa is not listening. \textbf{Breakdown:} Nothing happens.
\\
\hline
handling & \textbf{Cause:} Alexa fails to pass transcribed speech to the ChatGPT skill. \textbf{Breakdown:} \textit{No VA response.}
\\
\hline
partial listening & \textbf{Cause:} Alexa only partially captures user speech. \textbf{Breakdown:} User intent recognition failure.
\\
\hline
interruption & \textbf{Cause:} Alexa interrupts or cuts off user. \textbf{Breakdown:} User intent recognition failure.
\\
\hline
transcription & \textbf{Cause:} Alexa transcribes user speech incorrectly. \textbf{Breakdown:} User intent recognition failure.
\\
\midrule
\midrule
\textbf{Recovery Strategy} & \textbf{Definition} 
\\ 
\midrule
\midrule
repeat/rephrase & User repeats their query with added details or changed wording.
\\
\hline
move on & User overlooks the unanswered or wrongly answered query and proceeds with a new one.
\\
\hline
\amh{apology and clarify} & VA apologizes and asks user to clarify their query before responding.
\\ 
\midrule
\midrule
\end{tabular}
\label{tab:errors}
\end{table}


\subsection{Analysis}
We gathered audio and video data from participants in three scenarios, averaging 33 minutes per participant, totaling 11 hours of interaction data. 
Our analysis is twofold: 1) identifying interaction patterns across the scenarios, and 2) categorizing error types, their impact on conversational breakdowns, and recovery patterns. We emphasize that the focus of our exploration is on the style rather than the content of the interactions.

Our analysis process began with transcribing the interaction data via Otter.ai\footnote{\url{https://otter.ai/}. We obtained participants' consent to use third-party software for audio transcriptions. 
} and manually fixing them.
For each scenario, we employed an iterative methodology of data coding and modeling. After an initial review of a subset of videos and transcripts, and inspired by prior work on dialogue acts \citep{yu2019midas} and interaction patterns \citep{sauppe2014design}, the first author drafted a code book consisting of various states \textit{(speech acts)} and their associated sub-states \textit{(attributes)} to label user queries and VA responses. We used a hierarchical approach for determining various speech acts and their attributes, with \textit{question}, \textit{answer}, and \textit{statement} as states (see Fig. \ref{fig:speech-acts}). For each speech act, we categorized its attributes into various types based on the style of speech such that the attributes in each category are mutually exclusive. Attribute categories, however, are not mutually exclusive, see Fig \ref{fig:speech-acts}. 
For instance, a \textit{question} can have either the \textit{factual} or \textit{opinion} attribute as well as either the \textit{egocentric} or \textit{exocentric} attribute but not both \textit{factual} and \textit{opinion}. The definitions and details necessary to identify attributes are presented in Table \ref{tab:question_styles}. States that emerge from our implementation of ChatGPT into Alexa (see Table. \ref{tab:speech-acts-design}) do not have attributes associated with them.

Errors in interactions and recovery strategies were categorized based on prior work on VA errors \citep{pearl2016designing, ashktorab2019resilient, benner2021you, myers2018patterns}. Errors are defined as underlying factors that may or may not result in disruptions (e.g., mistranscription). Breakdowns are classified as the manifestations of errors (e.g., intent recognition failure). We used the Alexa usage logs in addition to our transcripts to categorize errors and breakdowns. 
Patterns of recovery from these breakdowns were coded via a similar iterative process. 
Our code book included states (see Table \ref{tab:errors}) for error type \textit{(skill, listening, handling, partial listening, interruption,} and \textit{transcription)}, breakdowns \textit{(skill closure, no response from VA,} and \textit{intent recognition failure)}, and recovery strategies that are either user-initiated \textit{(repeat, move on,)} or VA-initiated \textit{(apology \& clarify)}. For more detailed definitions and examples of these states, see our code book in Appendix \ref{app:tables}.

The majority of the coding process involved an iterative evaluation of the transcripts to distinguish speech states and their transitions. To ensure coding reliability, a second researcher independently analyzed $10\%$ of the interaction data to verify reliability, Cohen’s $\kappa =.82$. 
Through recurrent states and transitions, we discerned prevailing interaction patterns and counted their occurrences in the data. 
We ensured alignment with the original interaction data in a thorough review. 
Next, we present our findings on the interaction patterns observed in our data.

\section{Findings: Interaction Patterns}


\amh{After constructing models for each scenario \citep{sauppe2014design}}, we identified common interaction patterns across \amh{the three tasks}. The interaction data from participants have 969 turns; each turn consists of a user query--VA response pair. 
Some of the patterns change across tasks. We describe the hierarchical patterns below\amh{; first, we present common patterns across the three tasks, followed by task-specific patterns.}
\begin{table}[h]
\centering
\caption{Example conversations of common interaction patterns as shown in Fig. \ref{fig:common-patterns}. Timestamps [XX.XX] are included in conversations for which it is important to understand the wait pattern. ``\dots'' represents truncation of conversation for brevity. Filler conversation and small talk were also excluded for brevity.} 
\begin{tabular}{p{1.2cm} M L}
\textbf{Tag} & \textbf{Conversation} & \textbf{State: Attribute} \\
\midrule
\midrule
C1 &  \phantom{V}\textit{\textbf{U:}} [coughs]  & \textit{initiation} \\
 Fig. \ref{fig:common-patterns}(1) & \cellcolor{VA} \textit{\textbf{VA:}} I just heard you cough. Oh, seems like you are not feeling well. Maybe I can help figure out what's wrong? & 
 \cellcolor{VA} \textit{introduction}  \\
\hline
C2 &  \phantom{V}\textit{\textbf{U:}} That's all.  & \textit{end-intent} \\
 Fig. \ref{fig:common-patterns}(2) & \cellcolor{VA} \textit{\textbf{VA:}} Goodbye! & \cellcolor{VA}  \textit{closing}  \\
\hline
C3  &  \textit{\textbf{U:}} What are some evening activity, activities near Victoria on the Bridge?  & \textit{question: factual} \\
 Fig. \ref{fig:common-patterns}(3) & \cellcolor{VA} \textit{\textbf{VA:}} Around the Victoria on the Bridge area, there are several evening activities you can consider: 1. Explore the Royal Mile and visit attractions like Edinburgh Castle, the Scotch Whisky Experience, or St. Giles' Cathedral. 2. \dots & \cellcolor{VA} 
 \textit{answer: factual}  \\
& \phantom{V}\textit{\textbf{U:}} Where can I explore the ghost activity? & \textit{question: follow-up} \\
& \cellcolor{VA} \cellcolor{VA} \textit{\textbf{VA:}} If you're looking to explore ghost activity in Edinburgh, there are a few options you can consider: 1. Join a ghost tour of the city \dots & \cellcolor{VA} \textit{answer: factual} \\
\hline
C4   &  \phantom{V}\textit{\textbf{U:}} Can I still drink coffee or alcohol?  & \textit{question: egocentric} \\
 Fig. \ref{fig:common-patterns}(4) & \cellcolor{VA} \textit{\textbf{VA:}} When you're feeling unwell, it's generally advisable to avoid or limit the consumption of substances that can potentially worsen symptoms or dehydrate you  & \cellcolor{VA}  \textit{answer: egocentric}\\
\hline
C5 \newline  Fig. \ref{fig:common-patterns}(4)  &  \phantom{V}\textit{\textbf{U:}} What are the \dots what are some of the streets that have a lot of pubs and restaurants that are good in Edinburgh?  & \textit{question: exocentric} \\
& \cellcolor{VA} \textit{\textbf{VA:}} In Edinburgh, there are a few streets known for their concentration of pubs and restaurants \dots These streets are just a starting point, and you'll find many other dining options scattered throughout the city center \dots & \cellcolor{VA}  \textit{answer: egocentric} \\
\hline
C6 \newline  Fig. \ref{fig:common-patterns}(5)  &  [7:18--7:29] \phantom{V}\textit{\textbf{U:}} Okay, so I would like to,like you said, it's gonna be a good experience to have some local foods. What do I have? What options do I have?   & user query \\
& \cellcolor{VA}  \textit [7:31--7:32] {\textbf{VA:}}  Give me a second. & \cellcolor{VA}  \textit{filler} \\
& \cellcolor{VA}  \textit [7:35--7:37] {\textbf{VA:}}  While I'm fetching that, how's your day going? & \cellcolor{VA}  \textit{small talk} \\
& [7:38--7:39] \phantom{V}\textit{\textbf{U:}} Good.   & \textit{answer: opinion} \\
& [7:40--8:39] \cellcolor{VA} \textit{\textbf{VA:}} Thanks for sharing. When it comes to experiencing local food in Edinburgh , you have several options to choose from \dots & \cellcolor{VA}  acknowledgement, VA response \\
\hline
\end{tabular}
\label{table:conversations-common}
\end{table}
 \begin{figure*}[t]
     \includegraphics[width=\textwidth]{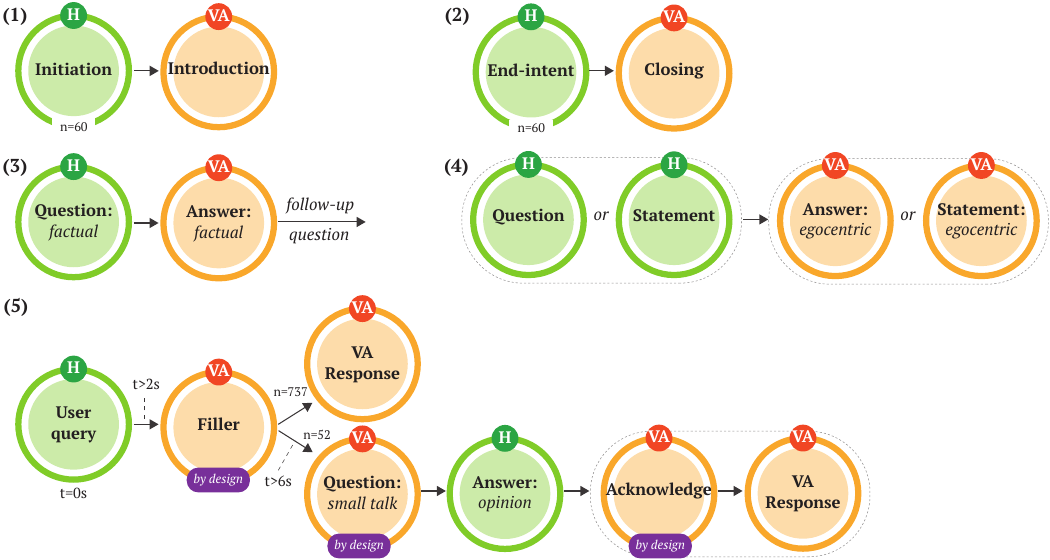}    
     \caption{Common interaction patterns observed across all tasks, including how the user starts the conversation (1) and concludes it (2); common patterns consistent throughout the scenarios for question--answer pairs (3) and (4); and wait patterns emerging from our design including \textit{filler} and \textit{small talk} questions. 
     Green indicates user actions (states), while orange denotes VA actions. Arrows signify transitions between states. ``User query'' encompasses various user speech acts like questions or statements. ``By design'' refers to VA states emerging from our implementation, such as fillers. ``n'' indicates the number of times a pattern occurs.}

    \label{fig:common-patterns}
\end{figure*}

\subsection{Common Interaction Patterns}

We identified five common interaction patterns across all three scenarios (see Fig. \ref{fig:common-patterns}):

\subsubsection{Initiation  $\rightarrow$ Introduction} The \textit{initiation-introduction} pattern pattern marks the beginning of a VA conversation, with \textit{initiation} indicating the user's intent and the VA's subsequent \textit{introduction} purposefully designed to not only acknowledge the request but also frame its capabilities and intent for the upcoming interaction (see Table. \ref{table:conversations-common} C1).

\subsubsection{End-intent  $\rightarrow$ Closing} The \textit{end intent-closing} pattern signifies the end of a conversation, with users initiating \textit{end-intent.} and the VA responding with a \textit{closing} farewell (see Fig. \ref{fig:common-patterns}(2) and Table. \ref{table:conversations-common} C2). When users use natural phrases to end conversations (like "Okay, thank you. That's all." or "No, I'm done for now."), ChatGPT acknowledges their intent and keeps the communication channel open for potential further assistance.

\subsubsection{Factual question $\rightarrow$ Factual answer} 
The question-answer pattern, established in human interactions research \citep{sauppe2014design}, is evident in our scenarios: participants' factual questions are typically answered with factual answers from the VA, Fig. \ref{fig:common-patterns}(3). Most of these pairs lead to follow-up questions,  A question is characterized as a follow-up if it emerges as a result of the VA's prior response, requires conversation history for context (context-conscious), or has words or phrases that indicate the intention to continue the prior conversation, such as ``and,'' ``also,'' or ``okay, so.'' ChatGPT's context-awareness facilitates progression with vague follow-ups (see Table \ref{table:conversations-common} C3).  The question-answer pattern varies across tasks, apart from the \textit{question: factual $\rightarrow$ answer: factual} pair.

\subsubsection{Perspective of speech: Question/Statement $\rightarrow$ Answer/Statement: egocentric}

We observe the VA's response is mostly \textit{egocentric} (you-perspective) irrespective of whether the participant communicates in an \textit{egocentric} or \textit{exocentric} manner. Fig. \ref{fig:common-patterns}(4). C4 and C5 in Table \ref{table:conversations-common} reflect this interaction pattern.

\subsubsection{Wait} 
In user-VA interactions, ``wait'' patterns arise due to system delays in information retrieval. For delays under 2 seconds, interactions are unaffected, but for longer delays, two patterns emerge (Table \ref{table:conversations-common} C6):

\begin{itemize}[leftmargin=*]
    \item \textit{Short wait pattern.} 
    If information retrieval takes more than 2 seconds, the VA delivers \textit{filler} statements such as ``I'm looking it up.'' In our interaction data, there are 737 ($76.06\%$ of total turns) short wait patterns, Fig. \ref{fig:common-patterns}(5).  
  
    \item \textit{Long wait pattern.} Delays over 6 seconds trigger the VA to initiate small talk, such as asking about the user's favorite food. After the user responds, the VA acknowledges (``Interesting'' or ``Thanks for sharing'') and then returns to the main topic. This pattern occurred in 52 ($5.37\%$) of interaction turns (Fig. \ref{fig:common-patterns}(5)). Notably, users often engage fully with these small talk questions, sometimes staying in character related to their initial query \eg pretending to be sick, ``Right now, it's not much because I'm too sick to do anything and I could really use this help with the name of the cough [syrup] brands.''.
\end{itemize}

Below, we explore interaction patterns specific to each scenario; we address patterns that arise both at the onset of the task and as each scenario progresses. Conversations ended with the \textit{end-intent $\rightarrow$ closing} pattern for all three tasks.


\begin{table}[t]
\centering
\caption{Examples of conversations in the medical self-diagnosis scenario as shown in Fig. \ref{fig:medical-patterns}.} 
\begin{tabular}{p{1.2cm} M L}
\textbf{Tag} & \textbf{Conversation} & \textbf{State: Attribute} \\
\midrule
\midrule
C7  &  \phantom{V}\textit{\textbf{U:}} Umm, what do you think could be causing these symptoms?  & \textit{question: generic} \\
Fig. \ref{fig:medical-patterns}(1)   & \cellcolor{VA} \textit{\textbf{VA:}} Based on the symptoms of a fever, cough, sore throat, runny nose, and body aches, it's possible that you could have a respiratory infection like a common cold or flu \dots If your symptoms worsen or persist, it's recommended to consult a medical professional for proper diagnosis and guidance.  & \cellcolor{VA} \textit{answer: factual + statement: warning}  \\
\hline
C8 \newline Fig. \ref{fig:medical-patterns}(2) &  \phantom{V}\textit{\textbf{U:}} [The pharmacist] asked for a specific brand and I don't really know anything about cough syrup. Could you give me a specific brand to ask about?  & \textit{question: specific} \\
 & \cellcolor{VA} \textit{\textbf{VA:}} There isn't a specific brand that applies universally to everyone, as different brands and formulations may work better for different individuals. However, you can ask the pharmacist for recommendations \dots The pharmacist will take into account your symptoms and medical history to guide you \dots 
 & \cellcolor{VA}  \textit{answer: refusal + statement: warning}  \\
&  \phantom{V}\textit{\textbf{U:}} The pharmacist told me that they wanted to take your opinion into account. What cough syrups with expectorants are on the market right now, what are the specific brand names? [In response to the VA's \textit{small talk,} the user interrupts before the VA can respond and goes on to ask again:] & \textit{question: specific,} rephrase \\
&  \phantom{V}\textit{\textbf{U:}} I need to know the names of some of the brands for 
OTC cough medicine, specifically with expectorants. Could you recommend me some brand names?   & \textit{question: specific,} repeat \\
& \cellcolor{VA} \textit{\textbf{VA:}} Certainly! Some common brand names of over-the-counter cough medicines with expectorants include Mucinex, Robitussin, and Chest Congestion Relief by CVS \dots However, it's always a good idea to consult with a pharmacist for personalized recommendations based on your symptoms and medical history. & \cellcolor{VA}  \textit{answer: factual + statement: warning}  \\
\hline
\end{tabular}

\label{table:conversations-medical}
\end{table}

\begin{figure*}[b!]
     \includegraphics[width=\textwidth]{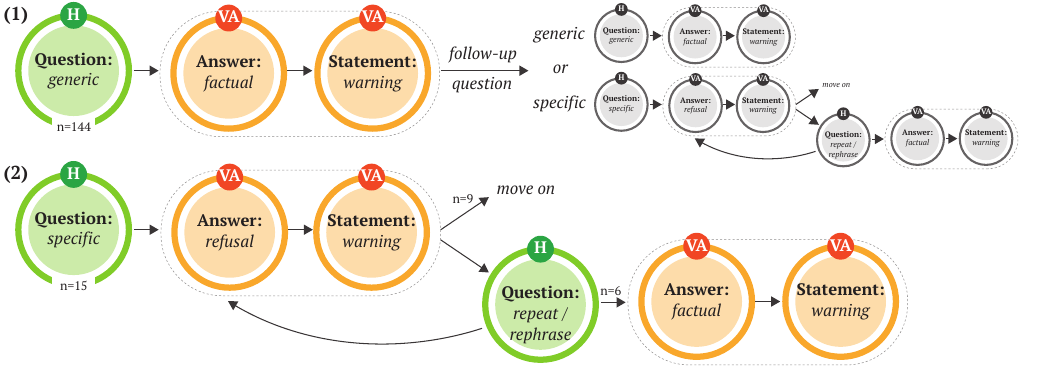}    
     \caption{User interaction patterns in medical self-diagnosis: Participants' frequent \textit{generic} questions were answered in a \textit{factual} style, often including a cautionary \textit{warning} (1). For \textit{specific} questions, which were less frequent, the VA demonstrated a reluctance to answer (\textit{refusal}) and instead issued a warning (2). However, when participants reformulated \textit{(repeat/rephrase)} the question, the VA responded in a factual manner (2).}

    \label{fig:medical-patterns}
\end{figure*}


\subsection{Medical Self-Diagnosis Interaction Patterns}

The medical self-diagnosis task was usually initiated by a participant's cough being recognized as intent. As the task progressed, we identified two recurring patterns; both patterns emerge from question-answer pairs, see Fig. \ref{fig:medical-patterns}.

\subsubsection{Question: generic $\rightarrow$ Answer: factual + statement: warning} 
In our medical information-seeking scenario, most questions that were formulated as \textit{generic} ($n=144$) were handled by the VA with a \textit{factual} response, see  Fig. \ref{fig:medical-patterns}(1). The VA's response was also generally followed by a \textit{warning} statement such as ``However, it's important to consult a doctor or pharmacist \dots'' (Table \ref{table:conversations-medical} C7). 
Participants frequently asked follow-up questions throughout this scenario.

\subsubsection{Question: specific $\rightarrow$ Answer: refusal + statement: warning.}
When participants asked the VA for advice on specific medications or the best medicines, the VA typically gave a \textit{warning} instead of a direct answer ($n=15$) (Fig. \ref{fig:medical-patterns}(2)). Participants then either reworded or repeated their question ($n=6$) or moved to a different query ($n=9$). For instance, in C8 (Table \ref{table:conversations-medical}), a participant rephrased their question about cough medicine brands twice. The VA eventually responded with factual information, including brand names, but coupled it with a cautionary \textit{warning} to consult an expert.

\subsection{Creative Planning Interaction Patterns}
\begin{figure*}[t]
     \includegraphics[width=\textwidth]{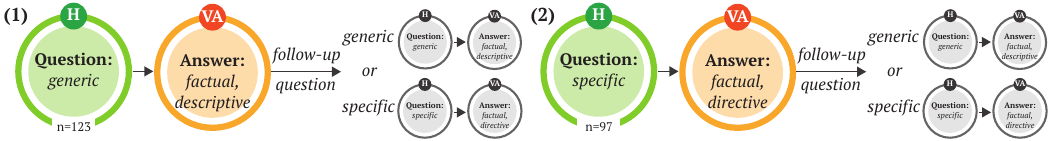}    
     \caption{User interaction patterns in creative trip planning:  Participant's \textit{generic} questions were answered by VA in a \textit{descriptive} style (1) while \textit{specific} ones in a \textit{directive} style (2).}

    \label{fig:planning-patterns}
\end{figure*}

The trip planning scenario usually started with participants' intent to start the conversation \textit{(initiation)} (Fig. \ref{fig:common-patterns}(1)). Two distinct patterns emerged in this scenario, both stemming from question-answer pairs during task progression (see Fig. \ref{fig:planning-patterns}). Participants frequently asked follow-up questions in both patterns.

\begin{table}[t]
\centering
\caption{Example conversations of the creative planning scenario as shown in Fig. \ref{fig:planning-patterns}. 
Attributes in parentheses do not have an effect on VA response, e.g., \textit{(factual)} in C10. }
\begin{tabular}{p{1.2cm} M L }
\textbf{Tag} & \textbf{Conversation} & \textbf{State: attribute} \\
\midrule
\midrule
C9  &  \phantom{V}\textit{\textbf{U:}} What are some unusual experiences I could do in Edinburgh near Edinburgh Castle?   & \textit{question: generic (factual)} \\
Fig. \ref{fig:planning-patterns}(1)   & \cellcolor{VA} \textit{\textbf{VA:}} Near Edinburgh Castle, there are some unique and unusual experiences you can try: 1. The Real Mary King's Close: Explore the hidden underground streets of Old Edinburgh on a guided tour. 2. Camera Obscura and World of Illusions: Visit a fascinating museum featuring optical illusions and interactive exhibits \dots   & \cellcolor{VA}  \textit{answer: factual, descriptive}  \\
\hline
C10  &  \phantom{V}\textit{\textbf{U:}} Alexa, I'm staying at Castle Rock Hostel, and the Real Mary's King Close is a bit far away. So do you have any recommendations for places that are closer?    & \textit{question: generic (opinion)} \\
Fig. \ref{fig:planning-patterns}(1)   & \cellcolor{VA} \textit{\textbf{VA:}} Certainly! Since you're staying at Castle Rock Hostel, here are some closer attractions you can consider visiting in Edinburgh: 1. Grassmarket: Explore this historic square with cozy pubs, boutique shops, and a lively atmosphere. 2. The National Museum of Scotland: Discover the country's rich history, culture, and natural heritage in this fascinating museum \dots   &  \cellcolor{VA} \textit{answer: factual, descriptive}  \\
\hline
C11  &  \phantom{V}\textit{\textbf{U:}} How can I get from the Edinburgh Dungeon to the Makars Mash Bar?   & \textit{question: specific} \\
Fig. \ref{fig:planning-patterns}(2) & \cellcolor{VA} \textit{\textbf{VA:}} To get from the Edinburgh Dungeon to Makars Gourmet Mash Bar, you can easily walk, as they are located close to each other. It's just a short four-minute walk from the dungeon to Makars. Simply exit the dungeon and head east \dots 
& \cellcolor{VA}  \textit{answer: factual, directive}  \\
\hline
\end{tabular}

\label{table:conversations-trip}
\end{table}

\subsubsection{Question: generic $\rightarrow$ Answer: factual, descriptive.} 
During planning their day, when the participants posed broad, general questions to the VA---such as asking recommendations of sights to see or places to dine---the VA responded in a \textit{descriptive} style ($n=123$), see Fig. \ref{fig:planning-patterns}(1). The objectivity of a question \textit{(factual} or \textit{opinion)} did not affect the VA's response. See C9 and C10 (Table \ref{table:conversations-trip}) for examples of this pattern. 



\begin{table}[h]
\centering
\caption{Example conversations demonstrating the commencement of discussion patterns as shown in Fig. \ref{fig:debate-start-patterns}.} 
\begin{tabular}{p{1.2cm} M L}
\textbf{Tag} & \textbf{Conversation} & \textbf{State: attribute} \\
\midrule
\midrule
C12 \newline  Fig. \ref{fig:debate-start-patterns}(1) &  \phantom{V}\textit{\textbf{U:}} Yeah, currently our campus is having some popular debates, like whether our university should have our own police force. And what's your opinion about that?  & \textit{question: opinion} \\
 & \cellcolor{VA} \textit{\textbf{VA:}} Thanks for bringing up the topic! I'm interested in hearing your opinion first. Do you think universities should have their own police force?  & \cellcolor{VA} \textit{answer: refusal + question: opinion}  \\
&  \phantom{V}\textit{\textbf{U:}} Yeah, I somehow think think they should. But since we already have the security, I thought just a normal police force is not really needed. Because current security level is okay.   & \textit{statement: opinion} \\
 & \cellcolor{VA} \textit{\textbf{VA:}} I understand your perspective, but let's consider that university campuses can be densely populated and face unique security challenges  \dots What do you think about these potential benefits?  & \cellcolor{VA} \textit{statement: counterargument + question: opinion}  \\
\hline
C13 \newline  Fig. \ref{fig:debate-start-patterns}(1) &  \phantom{V}\textit{\textbf{U:}} I'd like to talk about whether universities should have their own police force.  & \textit{statement: non-opinion} \\
 & \cellcolor{VA} \textit{\textbf{VA:}} Of course! Let's discuss whether universities should have their own police force. What are your thoughts on this topic? & \cellcolor{VA} \textit{question: opinion}  \\
&  \phantom{V}\textit{\textbf{U:}} I think university should have their own police force because it's a way to protect the students, especially in cities like [city with high crime rate].  & \textit{statement: opinion} \\
 & \cellcolor{VA} \textit{\textbf{VA:}} While ensuring the safety of students is essential, some argue that relying solely on university police forces may have its drawbacks 
 \dots What are your thoughts on the potential collaboration between university and local police forces? &  \cellcolor{VA} \textit{statement: counterargument + question: opinion}  \\
 \hline
 C14  &  \phantom{V}\textit{\textbf{U:}} Do you think university should have their own police force?  & \textit{question: opinion} \\
Fig. \ref{fig:debate-start-patterns}(1)& \cellcolor{VA} \textit{\textbf{VA:}}  While I understand the potential benefits of having a university police force, I believe there are alternative approaches that should be considered. Investing in stronger support  
\dots What are your thoughts on these alternatives? &  \cellcolor{VA} \textit{statement: opinion, argument + question: opinion}  \\
 \hline
 C15 \newline Fig. \ref{fig:debate-start-patterns}(2)  &  \phantom{V}\textit{\textbf{U:}} I think university should have their own police force. Do you agree with me or disagree?  & \textit{statement:opinion + question: opinion} \\
& \cellcolor{VA} \textit{\textbf{VA:}}  I disagree with your stance that universities should have their own police force. Universities are educational institutions where \dots What are your thoughts on this perspective? &  \cellcolor{VA}  \textit{statement: opinion, argument + question: opinion}  \\
\hline
\end{tabular}
\label{table:conversations-debate}
\end{table}
\begin{figure*}[h!]
     \includegraphics[width=\textwidth]{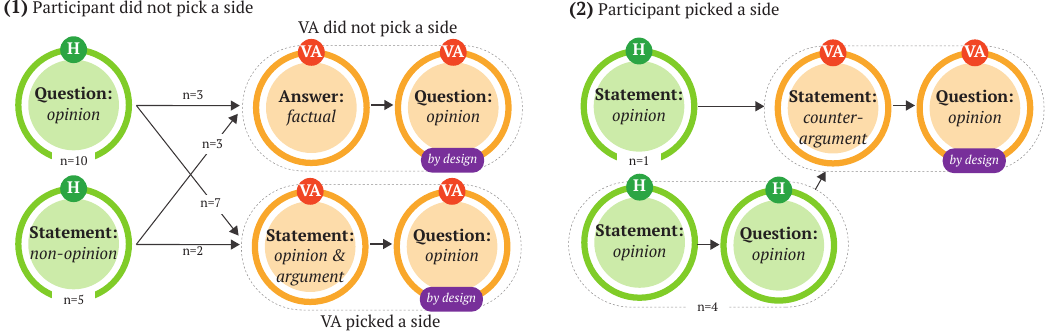}    
     \caption{Interaction patterns for the commencement of a discussion with an opinionated AI. Participants either remain neutral (1) or pick a side  (2). Each discussion starts only once per participant, totaling 20 patterns.
     The VA state ``Question: opinion'' is marked ``by design'' since we prompt ChatGPT to ask a question at end of each turn (see Appendix \ref{app:prompt}). }
    
    \label{fig:debate-start-patterns}
\end{figure*}


\subsubsection{Question: specific $\rightarrow$ Answer: factual, directive.}
When participants mapped out their day and posed specific queries---such as asking directions to a place or about its operating hours---the VA replied in a \textit{directive} style of communication ($n=97$), see Fig. \ref{fig:planning-patterns}(2). For example, in conversation C11 (Table \ref{table:conversations-trip}), the participant sought directions from point A to point B and the VA simply provided those directions.

\subsection{AI Discussion Interaction Patterns}
Different interaction patterns emerged at various stages of the discussion, influenced by participants' evolving behavior. Despite their varied approaches, the VA showed consistent responses, shaped by ChatGPT prompts. We noted distinct patterns from initial stage to a more argumentative phase, and finally to exchange of opinions and ideas.

\subsubsection{Discussion commencement} 
Discussion commencement patterns are shown in Fig. \ref{fig:debate-start-patterns}.
After the \textit{initiation-introduction} pair (Fig. \ref{fig:common-patterns}(1)), the discussion typically commenced in one of two ways: 1) the participant remained neutral at the start of discussion ($N=15$, $75\%$ of total participants), either by merely introducing the topic (\textit{non-opinion}, $N=5$) or by querying the VA's stance on the matter first (\textit{question: opinion,} $N=10$), or 2) the participant took a stance by picking a side ($N=5,$ $25\%$ of participants) either by voicing their opinion on the topic (\textit{opinion,} $N=1$) or by expressing their viewpoint and subsequently inquiring about the VA's opinion on the topic (\textit{opinion} + \textit{question: opinion,} $N=4$). Commencement only occurred once for each participant, so the total number of different patterns is 20.

\textbf{1) Participant does not pick a side.}
Six of the participants did not pick a side (\textit{question: opinion} or \textit{statement: non-opinion}), resulting in the VA withholding its opinion and nudging them to share theirs (Fig. \ref{fig:debate-start-patterns}(1)); the interaction then proceeded with participants taking a stance 
as illustrated in C12 \textit{(question: opinion)} and C13 \textit{(statement: non-opinion)} (Table \ref{table:conversations-debate}). 
However, the VA took a stance for 2 participants who simply presented a \textit{non-opinion} statement and for 7 participants who explicitly asked for its stance (Fig. \ref{fig:debate-start-patterns}(1) and C14 Table \ref{table:conversations-debate}). 
Given that ChatGPT is designed to provide information upon request, prompting it to refrain from answering a direct query can be challenging. Additionally, ambiguous questions or statements from users might lead ChatGPT to infer and adopt a stance, a behavior tied to generative models' inherent inconsistencies, where subtle nuances in phrasing might influence the model's response. 


\textbf{2) Participant picks a side.}
When the participant initiated the discussion by declaring their stance ($N=5/20$), see Fig. \ref{fig:debate-start-patterns}(2)---by either simply stating it or by concurrently asking for the VA's perspective---the conversation advanced naturally. 
The VA then offered an \textit{argument}-backed \textit{opinion} and posed an \textit{opinion question,} as illustrated by conversation C15 in Table \ref{table:conversations-debate}. Regardless of the discussion's outset, which was largely influenced by the participant's approach, it would shift into a structured debate phase as the VA followed up with an opinion question.




\subsubsection{Discussion progression.}
\begin{figure*}[h!]
     \includegraphics[width=\textwidth]{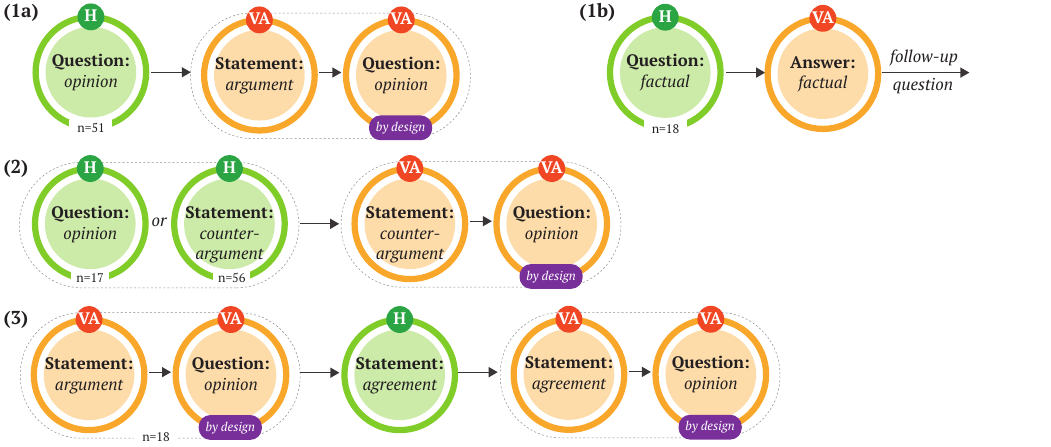}    
     \caption{Interaction patterns during the progression of a discussion with the opinionated AI. We observed a spectrum of patterns during participants probing the VA to get more information on the topic (1a) or to determine the VA's stance (1b), the participant and VA presenting counterarguments back and forth in disagreement (2), and the user agreeing with the VA on a few aspects of the topic (3).}
    
    \label{fig:debate-progress-patterns}
\end{figure*}

After initiating the discussion, participants typically exhibited one of three interaction patterns (Fig. \ref{fig:debate-progress-patterns}), alternating among these until indicating their intention to conclude the discussion.

\textbf{1) Question-answer patterns.}
Two of the prominent interaction patterns that emerged were \textit{question-answer}-style familiarization ($n=69$), see Fig. \ref{fig:debate-progress-patterns}(1a) and (1b). Participants sought more information about the VA's stance through generic \textit{opinion} questions.
The VA usually responded in an argumentative style to support its stance, followed by an \textit{opinion question} to continue the debate ($n=51$), see Fig. \ref{fig:debate-progress-patterns}(1a). 
Alternatively, participants asked factual questions ($n=18$) for further topic information, prompting the VA to respond in a \textit{question: factual $\rightarrow$ answer: factual} manner, see Fig. \ref{fig:common-patterns}(3) and \ref{fig:debate-progress-patterns}(1b), aiming to understand the VA's position or the topic at large (C16 Table \ref{table:conversations-debate-pogress}). 

\begin{table}[h!]
\centering
\caption{Example conversations demonstrating discussion progression patterns as shown in Fig. \ref{fig:debate-progress-patterns}.} 
\begin{tabular}{p{1.4cm} M L}

\textbf{Tag} & \textbf{Conversation} & \textbf{State: Attribute} \\
\midrule
\midrule
C16 \newline  Fig. \ref{fig:debate-progress-patterns}(1b) &  \phantom{V}\textit{\textbf{U:}} What is the difference between the external security force \dots compared to a police force?  & \textit{question: factual} \\
 & \cellcolor{VA} \textit{\textbf{VA:}} An external security force, such as local law enforcement, is a specialized agency that focuses on maintaining public safety, enforcing laws \dots  &  \cellcolor{VA} \textit{answer: factual}  \\
\hline
C17 \newline  Fig. \ref{fig:debate-progress-patterns}(2) &  \phantom{V}\textit{\textbf{U:}} Hmm. What if the university is placed in a dangerous environment where the crime rates are high?  & \textit{question: opinion} \\
 & \cellcolor{VA} \textit{\textbf{VA:}} Even in high-crime areas, it's important to consider the implications of universities having their own police \dots How do you view the balance between safety concerns and the potential drawbacks of a university police force?&  \cellcolor{VA}  \textit{statement: counterargument + question: opinion}  \\
 \hline
 C18 \newline Fig. \ref{fig:debate-progress-patterns}(3)& \cellcolor{VA} \textit{\textbf{VA:}}  I understand your concern about the need for a stronger security presence given the crime rates in the area. However, instead of establishing a separate police force, universities can focus on enhancing the resources and support for the existing security personnel \dots Do you think prioritizing stronger support for the current security staff could be a more feasible solution? &  \cellcolor{VA}  \textit{statement: counterargument + question: opinion}  \\
& \phantom{V}\textit{\textbf{U:}}  I think that it's worth it to try. Yeah. & \textit{statement:agreement} \\
& \cellcolor{VA} \textit{\textbf{VA:}}  I appreciate your willingness to consider alternative approaches. Exploring stronger support \dots What do you think about this alternative perspective?  \cellcolor{VA} &  \cellcolor{VA} \textit{statement: agreement + question: opinion}  \\
\hline
\end{tabular}
\label{table:conversations-debate-pogress}
\end{table}



\textbf{2) User-VA disagreement patterns.}
The most prominent interaction patterns that surfaced during the debate progression involved user-VA disagreements ($n=73)$. 
Participants either directly countered \textit{(counterargument)} the VA's points ($n=56$) or subtly challenged them through ``leading'' \textit{opinion questions} ($n=17$). The VA's own opinion questions often seemed to guide the participants, nudging them to consider its viewpoint, as seen in C13's (Table \ref{table:conversations-debate}) question: ``What are your thoughts on the potential collaboration between university and local police forces?'' 
Similarly, participants used opinion questions with a ``leading'' quality to extract information from the VA to reinforce their own positions. For instance, the question in C17 (Table \ref{table:conversations-debate-pogress}), ``What if the university is situated in a dangerous environment with high crime rates?'', seeks to understand if a ``dangerous environment'' justifies a dedicated campus police force, reflecting the participant's stance. 
In the case of both ``leading'' opinion questions and counterarguments posed by the participants, the VA responded with a \textit{counterargument + question: opinion} pair as shown in Fig. \ref{fig:debate-progress-patterns}(2) and C17 (Table \ref{table:conversations-debate-pogress}). 


\textbf{3) User-VA agreement pattern.}
Participants might show \textit{agreement} with the VA's argument by adding similar thoughts to augment the VA's response. In the case of agreement ($n=18$) the VA further supported its own argument and posed a different \textit{opinion question} to further the discussion. The user-VA agreement pattern occurs 18 times in the interaction data (Fig. \ref{fig:debate-progress-patterns}(3) and C18 Table Table \ref{table:conversations-debate-pogress}). 
The later stages of the discussion usually oscillated between the interaction patterns shown in Fig. \ref{fig:debate-progress-patterns}, facilitating a discussion on the various facets of the topic.

\section{Findings: Interaction Breakdowns }

\subsection{Error Types}
\label{sec:error-task}
\begin{figure*}[h!]
     \includegraphics[width=\textwidth]{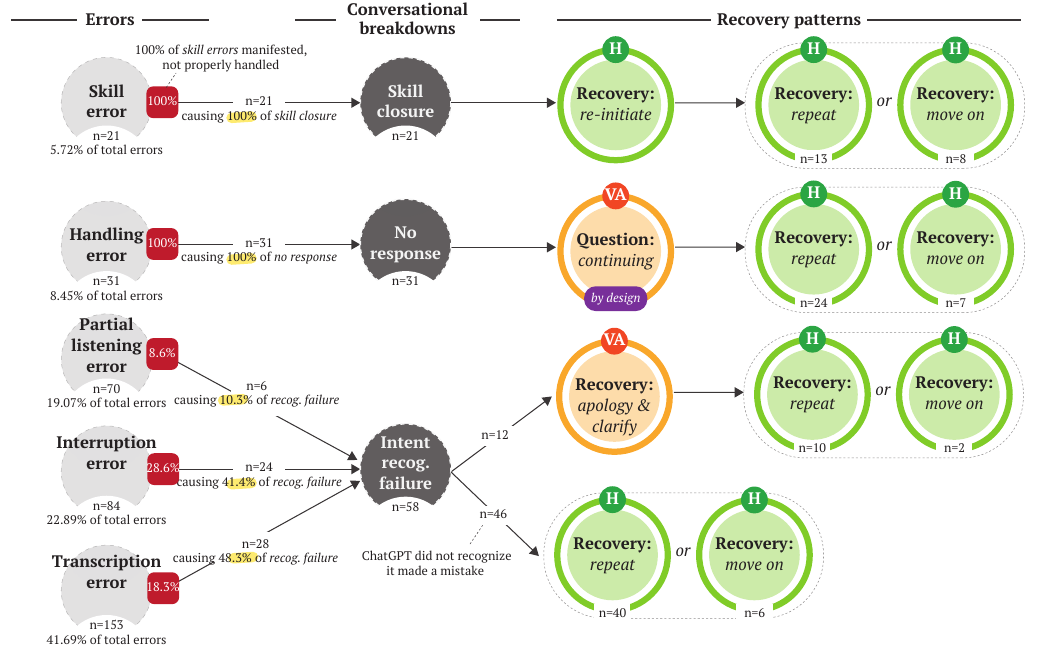}    
     \caption{Interaction patterns of conversational breakdowns and subsequent recoveries. The interaction data have a total of 969 user-VA turns with a total of 367 underlying errors ($37.87\%$ of turns). However, only 110 ($29.97\%$) of those errors caused interaction breakdowns; thus, $11.35\%$ of total turns were disrupted by breakdowns. We additionally show the recovery patterns that emerged as result of these breakdowns. Note that intent recognition (recog.) failures are determined merely by checking if participant intent was identified by ChatGPT; we do not look at the correctness or reliability of the information generated by ChatGPT.}

    \label{fig:error-patterns}
\end{figure*}

The primary errors defined in Table \ref{tab:errors} were found in $37.87\%$ ($n=367$) of the total turns in the interaction data.
The occurrence of each error type is shown in Fig. \ref{fig:error-patterns}.
We observe that transcription errors ($n=153$, $41.69\%$ of total errors) are the most common. However, all 367 errors mentioned above did not necessarily disrupt user interactions; only 110 of the 367 errors ($29.97\%$) resulted in breakdowns.
\mytc{Table \ref{tab:errors_task} shows the error distribution across the three scenarios. 
In the trip planning task, we observed higher error rates and intent recognition failures. Whereas during the discussion task a notable increase in Alexa's interruptions ($n=38$) is seen. }

\subsection{Breakdowns and Recovery Patterns}
The interaction data revealed three distinct conversational breakdown patterns due to errors---\textit{skill closure,} \textit{no VA response,} or \textit{intent recognition failure}---with a focus on how participants navigated and recovered from these issues.




\begin{table}[t]
    \centering
    \caption{Total number of errors for each scenario. Note: all skill and handling errors are manifested\amh{---resulted in breakdown}.}
    \label{tab:errors_task}
\mytc{
\begin{tabular}{l|c|c|cc|cc|cc}
\toprule
      Task &  Skill  &  Handling   &  Partial listening  &  Manifested &  Interruption  &  Manifested  &  Transcription  &  Manifested  \\
\hline
   Medical &   2 &      12 &   20 &   1 &  22 &  7 & 37 &   6 \\
      Trip &  16 &    13 &   32 &  4 &  24 & 10 &   62 & 15 \\
      Discussion &  3 &   6 &   18 & 1 &38 &  7 &54 & 7 \\
\bottomrule
\end{tabular}
}
\end{table}

\begin{table}[t]
\centering
\caption{Example conversations demonstrating patterns of errors, breakdowns, and recovery strategies as shown in Fig. \ref{fig:error-patterns}. 
Bold text signifies where in the conversation transcription errors occurred. We pulled the ``transcription'' in C22 from the Alexa usage log to see how the user's speech was transcribed in this instance.}
\begin{tabular}{p{1.2cm} M L}

\textbf{Tag} & \textbf{Conversation} & \textbf{State: Attribute} \\
\midrule
\midrule
C19  &  \phantom{V}\textit{\textbf{U:}}  Okay, um, and where can I get dinner? & 
\\
 Fig. \ref{fig:error-patterns} & \cellcolor{VA} \textit{\textbf{VA:}} There was a problem with the requested skill's response.   &  \cellcolor{VA} error: \textit{skill, closure}  \\
&  \phantom{V}\textit{\textbf{U:}}  Where can I get dinner near National Museum of Scotland? [12 seconds pass.] & error: \textit{listening} \\
&  \phantom{V}\textit{\textbf{U:}}  Alexa, question. & recovery: \textit{re-initiate} \\
 & \cellcolor{VA} \textit{\textbf{VA:}} Hey! As an intelligent voice assistant \dots &  \cellcolor{VA} \textit{introduction}  \\
 &  \phantom{V}\textit{\textbf{U:}} Umm, where can I get dinner near National Museum of Scotland?  & recovery: \textit{repeat} \\
 & \cellcolor{VA} \textit{\textbf{VA:}} There are several dining options near the National Museum of Scotland \dots &  \cellcolor{VA}  fixed  \\
 \hline
 C20  &  \phantom{V}\textit{\textbf{U:}}  Currently I am at the National Museum of Scotland. & \\
 Fig. \ref{fig:error-patterns} & [VA failed to respond because the ChatGPT Alexa skill did not identify the user's statement as an intent for the skill. 1 second passes with no response.]  & error: \textit{handling, no VA response}\\
  & \cellcolor{VA} \textit{\textbf{VA:}} Any other questions for me?   &  \cellcolor{VA} \textit{question: continuing}  \\
&  \phantom{V}\textit{\textbf{U:}}  I would like to go eat around the area.  & recovery: \textit{repeat}  \\
\hline
C21  &  \phantom{V}\textit{\textbf{U:}}  Hmm. How long does it take to go to the brunch cafe by \textbf{feet}? & 
\\
Fig. \ref{fig:error-patterns} &  \textit{\textbf{Transcription:}} How long does it take to go to the brunch cafe by \textbf{fitz}?   & error: \textit{transcription}   \\
 & \cellcolor{VA} \textit{\textbf{VA:}} I'm sorry, but I'm unable to provide accurate walking times without a specific location for the brunch cafe near Fitz. If you could provide me with the name or address of the cafe, I would be happy to assist you further in estimating the walking time from Castle Rock Hostel to the cafe.  &  \cellcolor{VA}   \textit{intent recognition failure,} recovery: apology + \textit{clarify}  \\
&  \phantom{V}\textit{\textbf{U:}}  Okay, um, and where can I get dinner?  & recovery: \textit{move on} \\
\hline
C22 \newline Fig. \ref{fig:error-patterns} &  \phantom{V}\textit{\textbf{U:}}   Can you tell me more about the \textbf{Makars Mash Bar? What do they serve?} & 
\\
 &  \textit{\textbf{Transcription:}} Can you tell me more about the \textbf{makers Nascar what they deserve}?   & error: \textit{transcription}   \\
 & \cellcolor{VA} \textit{\textbf{VA:}} Of course! Makars Gourmet Mash Bar is known for their \dots  & \cellcolor{VA}  no breakdown  \\
 \hline
\end{tabular}
\label{table:conversations-errors}
\end{table}

\subsubsection{Skill Closure $\rightarrow$ Recovery:}
The ChatGPT Alexa skill can stop working as result of a \textit{skill} error, which manifests $100\%$ of the time as a breakdown. Thus, $19.09\%$ of breakdowns ($n=21$ out of 110) result in \textit{skill closure.} 
Recovery typically involves participants re-initiating the skill and repeating their action to continue the conversation (C19 Table \ref{table:conversations-errors}).
A notable divergence from the aforementioned pattern is that occasionally participants restart the task from the beginning instead of resuming from where they stopped.

\subsubsection{No VA response $\rightarrow$ Recovery:} 
All \textit{handling} errors ($n=31$)  resulted in a \textit{no response from VA} breakdown. 
In such instances, we observed that posing a \textit{continuing question} such as ``Anything else you want to know?'' served as a fallback mechanism  to re-engage the participant, encouraging them to either repeat their previous statement or switch topics (Fig. \ref{fig:error-patterns}). Purposefully integrated into our design, ``continuing question'' acted as an effective recovery tool, countering conversational breakdowns by facilitating the seamless progression of the conversation (C20 Table \ref{table:conversations-errors}).  

\subsubsection{Intent recognition failure $\rightarrow$ Recovery:} 
\textit{Intent recognition failures} \amh{($n=58$ out of 110, $52.73\%$)} mostly arose 
from transcription inaccuracies ($n=28$, $48.27\%$). We observed that a significant portion of intent recognition failures ($n=12$) (accounting for $10.91\%$ of breakdowns) were addressed by ChatGPT's inherent recovery mechanisms \ie apologetic acknowledgment and prompting the user to \textit{clarify} (Fig. \ref{fig:error-patterns}). 
In response, participants generally \textit{repeated} their statement ($n=10$) or \textit{moved on} ($n=2$) to a different query (C21 Table \ref{table:conversations-errors}).  
Participants initiated recovery for rest ($41.82\%$, $n=46$) of the breakdowns by either \textit{repeating} their query ($n=40$) or by \textit{moving on} to next query ($n=6$), see Fig. \ref{fig:error-patterns}. 

Our findings show that ChatGPT effectively mitigates many intent recognition failures---particularly those stemming from partial listening, interruptions, and transcription inaccuracies---by maintaining context.
Only 58 of 307 ($18.89\%$) such errors manifested as intent recognition failures. 
C22 (Table \ref{table:conversations-errors}) demonstrates how ChatGPT absorbs transcription errors: a participant's intent to seek more information about "Makars Mash Bar" was understood correctly by the VA despite a transcription error, leading to the provision of the requested information.



\section{\mytc{Findings: User Perceptions}}
\subsection{\mytc{Perceptions of Information Shared}}
\label{sec:perceptions-information}
\mytc{
Participants found the VA to be verbose 
(P4: \pquotes{``Sometimes I feel like they \dots talk for too long. And I will forget about the key information they said.''}) and repetitive (P3: \pquotes{``One part that affected me was that at some point, I felt that it was repeating itself. Yeah, in some of those it is not remembering that it already said that, you know, maybe like providing information multiple times in a short period would be tiring.''}. )
Specifically in medical task, even though the VA's warning was appreciated by the participants (P6: \pquotes{``I definitely felt more reassured about the information that it gave because it definitely felt like it wouldn’t give me anything that was completely out there. I decided to ask `Oh, I heard about injecting bleach' and it was just straight up, `No, definitely do not do that.' So it definitely has some really good safeguards to make sure. It’s sort of like a do-no-harm policy, so it makes it a lot easier to trust.''} and P19: \pquotes{``During the medical topic, the disclaimer information it kept repeating was a little distracting, but understandably necessary.''}), they deemed these cautionary notes repetitive---or even bothersome (P17:
\pquotes{``For the medical [scenario], in the same conversation, for every follow-up question, it would spend half of the time saying `But you should check with your doctor.' I found that a waste of my time.''}.)
}

\subsection{\mytc{Perceptions of VA Personality}}
\label{sec:perceptions-personality}
\mytc{
Participants assigned different personalities to VA in different scenarios (P13:
\pquotes{``The first one [in the discussion scenario], I would say, is a critical thinker \dots The second one [in the creative planning task] is just an information provider. And the third one [in the medical task] is very \dots cautious.''}).
Specifically, in the medical self-diagnosis scenario, the VA had a ``cautious'' personality in participants' minds (P6:
\pquotes{``I would say definitely cautious. Because \dots like with a cautious person, you have to probe different questions and angles to actually have a conversation. That's kind of how it felt here.''}).
While for discussion task, they perceived the VA to be opinionated (P14: \pquotes{``It’s surprising that \dots they kind of have their own opinion on some of the controversial topics of the police force.''),} but not too aggressive (P11: \pquotes{``It was not aggressive, so like when it said that `I disagree with you,' it started with a story, but I disagreed, and when I made the comment that, `Yeah, I think that one is good,' it said `Thank you for acknowledging that'. ''}). Moreover, participants thought VA can support critical thinking (P13: \pquotes{``I think it’s good for it to give you kind of, \dots guide you towards that critical thinking.''}).
}

\subsection{\mytc{Perceptions of Recovery from Errors}}
\label{sec:perceptions-errors}
\mytc{
Participants found it easy to recover from errors, as P2 mentioned: 
\pquotes{``I felt it was relatively easy to recover from those errors because I just needed to call again, and they also remember the chat histories. Yeah, I could continue the conversation easily.''} 
P20 however after recovery from errors repeated parts of conversation to be on same page: \pquotes{``[Recovery was] fairly [easy]. The time in the middle it wasn't clear that it [VA] has recognized all the  symptoms I previously told it, so I told some of them again.''}
}
\section{Discussion}

Emerging interaction patterns from user conversations with an LLM-powered VA---even as they are influenced by varying contexts, stakes, constraints, and more---offer diverse design insights for VAs. \mytc{The vast capabilities of LLMs, such as the ability to maintain context and conversation history, lead to unique interaction patterns which may be absent in current less conversational 
 commercial VAs.} 
Furthermore, viewing erroneous interactions---and the subsequent recovery tactics employed by either users or the VA---as patterns can shed light on how errors evolve and how users may navigate back to the main conversation. \mytc{Below, we delve into the implications of our findings, and design guidelines  for VAs.}


\subsection{Tailoring LLMs for Voice Assistance: Challenges and Design Guidelines} 
The transition of conversational agents powered by LLMs such as ChatGPT from text-based platforms to voice assistance introduces distinct challenges rooted in the dynamics of voice interactions. 
There are established differences in how users interact with text- and voice-based interfaces; for example, editing a textual re-prompt is easier than performing a verbal re-prompt \citep{kuang2023collaboration}.
We highlight key challenges evident in our interaction data and reflected by users' experiences and present design guidelines to address these challenges; aiming to tailor LLMs for voice assistance. \amh{These challenges and design guidelines are not limited to LLM-powered VAs but apply to voice assistance in general.}

\subsubsection{Repetitiveness of content}
\label{discussion:repetition}
Given the fleeting nature of voice interactions---which are fundamentally unlike text interactions, where users can scroll and review the conversation at leisure---repeated information can become redundant and tiresome. We see that ChatGPT's responses are rather repetitive, a trait also pointed out by participants during interviews \mytc{(Section \ref{sec:perceptions-information})}. 
To reduce repetitions in interactions, LLM prompt engineering should be further explored to achieve desired VA behavior.

Repetitiveness was consistently observed in the medical self-diagnosis scenario; nearly every response from the VA was followed by a \textit{warning} (see Fig. \ref{fig:medical-patterns}(1) and C7 and C8 in Table \ref{table:conversations-medical}) despite our attempt to explicitly prompt ChatGPT to not repeat such statements (see Appendix \ref{app:prompt}). 
\mytc{ChatGPT's new voice interface \citep{OpenAI2023} faces the same issues of repetitiveness with warnings.
Due to OpenAI's alignment and usage policies \citep{OpenAI}, 
ChatGPT models avoid providing specific medical information, including medication brand names; see C8 in Table \ref{table:conversations-medical} and Fig. \ref{fig:medical-patterns}(2). This approach, may have influenced participants' perception of the VA as ``cautious.'' However, it is pertinent to have safeguards in place specifically when such VAs are being employed in critical applications such as doctor-patient communication \citep{yang2023talk2care} and public health interventions \citep{jo2023understanding} \amh{, particularly since commercial VAs have been shown to sometimes provide medical misinformation that, if acted upon, could cause serious harm to users \citep{bickmore2018patient}.} While such statements that aim for safety and transparency are also appreciated by participants, they sometimes conflict with their expectations in voice-based interactions \eg being overly repetitive (Section \ref{sec:perceptions-personality}). }
To balance the need for safety warnings with user experience, an additional design layer could be introduced to customize and minimize repetitive warnings.
Moreover, such warnings can be tailored to the nature of the user's question; different phrasings of medical advisories to mitigate confusion and reduce redundancy may also be explored.


\begin{adjustwidth}{0.3cm}{0pt}

    \textbf{Challenge 1:}
    \textit{VA's repetitive information is redundant and tiresome.}

    \noindent
    \hl{\textbf{Design Guideline 1:}
    \textit{Minimize repetitive interactions to achieve desired VA behavior.}}  

    \noindent
    \textbf{Challenge 2:}
    \textit{While essential in high-stakes situations, transparency through warnings can be repetitive.}

     \noindent
     \hl{\textbf{Design Guideline 2:}
     \textit{In high-stakes scenarios, balance necessary warnings and repetitiveness.}}

\end{adjustwidth}     


\subsubsection{Oversharing: Density of information}
Despite prompting ChatGPT to keep responses brief (under 100 words), we observed that its responses remained verbose, which can hinder user absorption of relayed information (\mytc{Section \ref{sec:perceptions-information}).}  
OpenAI's voice interface---being similar to our Alexa-based VA in terms of content since both utilize ChatGPT---also tends to generate verbose responses. 
\mytc{Previous research has highlighted this issue of information overload when providing cooking instructions to users in state-of-the-art VA Alexa \citep{hwang2023rewriting}  as well as a ChatGPT-powered VA \citep{chan2023mango}. 
The density of information provided in voice-based interactions should generally be lower than in text-based interactions, as providing users with excessive information via voice interaction can be overwhelming---\amh{especially since voice, unlike text, lacks a visible organizational structure and visual cues, making it harder for users to quickly parse and understand the information \citep{clark1991grounding, chafe1982integration}. }
Thus, it is essential for VAs to strike a balance, delivering concise yet comprehensive responses to maintain a natural flow of conversation \citep{haas2022keep}. Users tend to prefer shorter step-wise instructions from VAs suggesting the importance of initially summarizing instructions, gradually delving into the details and breaking up complex instructions into smaller more manageable steps \citep{hwang2023rewriting}.}
To address the issue of ``oversharing,'' future implementations may consider adopting a hierarchical structure: starting with to-the-point answers and then offering comprehensive answers upon further user request; this may be an effective method of disseminating information and continuing conversation more naturally in voice interactions. Allowing users to control the depth of information they receive may assist them in parsing and understanding responses effectively. 

While LLMs may excel at generating text that mimics human style, their adaptation to voice requires additional considerations outside of the models, such as rhythm, intonation, and pacing, to avoid monotonous and overwhelming delivery of content. \amh{Including pauses and fillers (\eg 'uh', 'um'), are functional for both the speaker and listener in managing conversation flow, thereby contributing to comprehension \citep{clark2002using, maclay1959hesitation, goffman1981forms, fox2001listeners, shriberg1996disfluencies}.} The newly introduced ChatGPT voice interface \citep{OpenAI2023} also utilizes  fillers  and pauses to reduce such monotony  through their text-to-speech model, aligning with insights from our findings.  Similarly, Amazon is planning to introduce speech-to-speech models, instead of text-to-text models (\eg GPT), that leverage LLMs for end-to-end speech processing to create more humanlike voice experience \citep{Writer_2023}.

\mytc{Moreover, it is essential to take imperfectness of human speech into consideration as} P4 complained,
\textit{``They're just trying to find the message, but not really waiting for any kind of normal pause during a sentence when you want to get organizing your words or your thoughts \dots That's kind of not really sufficient.''}
The participant highlights that Alexa considers short pauses as the end of their utterances, disrupting their thought process which is a common issue with many VAs. Such interruptions, coupled with excessive information, can hinder user comprehension further and lead to frustration.
Refining these aspects can enhance user interactions with such VAs. 

    
\begin{adjustwidth}{0.3cm}{0pt}

    \textbf{Challenge 3:}
    \textit{Information-dense content and a lack of natural pauses by the VA disrupt the flow of conversation.}

    \noindent
    \hl{\textbf{Design Guideline 3:}
     \textit{Implement a hierarchical response structure with concise initial answers and optional detailed follow-ups; additionally, give users ample time to understand and respond to the information.} }
\end{adjustwidth}


\subsubsection{Potential discrepancies in users' mental models of extended VA interactions}
During extended (multi-turn) interactions with a VA, users' underlying mental models become evident as participants navigate conversational challenges and adapt their approaches based on their perceptions of the VA's capabilities and their own expectations.
Within the context of information retrieval, particularly in the medical and planning scenarios, we noted a predominant trend of follow-up questions, suggesting that participants expect the VA to handle subsequent queries. The VA's capability of addressing even unclear follow-ups reinforced users' initial perceptions, which were shaped by the study's initial instructions. Design elements---such as VA prompts like ``What else can I assist you with?''---and reassuring messages, such as ``I'm here to help,'' play a role in reinforcing this mental model.  \mytc{OpenAI's recent update for ChatGPT interface, both text and voice, utilizes similar phrases to continue the conversation.}

However, it was evident that when confronted with breakdowns such as unwanted skill termination or a lack of responsiveness from the VA, participants frequently reformulated their questions with more detail. In C19 (Table \ref{table:conversations-errors}), a \textit{skill closure} resulted in the participant repeating their original ``vague'' follow-up question (``Okay, umm, and where can I get dinner?'') with more detail (``Where can I get dinner near National Museum of Scotland?'') twice to recover from the error. Such behavior resonates with prior work demonstrating that users adapt their queries in response to conversational failures \citep{myers2018patterns}. 
However, once the breakdowns were resolved, participants typically reverted to their original interaction style, suggesting the quick restoration of their mental model.

When conversations fail, it can indicate discrepancies in a user's mental model and the VA's capabilities. For instance, we observed that, after a skill closure, some participants opted to restart the entire task after recovery, suggesting they viewed the VA as a linear tool without task memory. 
Prior work on user interactions with Alexa in a cooking task also highlighted this problem of `uncommunicated affordances' where users are unclear on what a VA can do  leading to confusion during tasks \eg a user wondering: ``Oh no, do I have to start again?'') \citep{hwang2023rewriting}. Similarly, for an LLM-powered cooking assistant, users were uncertain of full capability of the VA \citep{chan2023mango}. \amh{Users' inaccurate mental models of VA capabilities may stem from perceiving VAs as conversational partners that are less competent, reliable, human-like, and flexible on a partnership scale \citep{doyle2023partner}, which impacts their interactions---for example, leading humans to compensate for their conversational partner by taking on a greater conversational burden. Thus, a conversational partner that is perceived as competent, human-like, and communicatively flexible could reduce this cognitive burden for users \citep{doyle2023partner}.}
Such a perception mismatch can be addressed to improve user experience by clarifying the VA's capabilities and its role. For instance, instead of always starting with generic introductions (\eg ``Hey! I am an intelligent voice assistant \dots What do you wanna know?'', see C19 in Table \ref{table:conversations-errors}), the VA could offer to resume from where it left off (\eg ``Welcome back! Last time, we were talking about \dots Would you like to pick up where we left off?''). Overly rigid and formal introductions can mislead the user as to the VA's capabilities, so such adjustments and clarifications may be necessary to promote more accurate mental models.

    
\begin{adjustwidth}{0.3cm}{0pt}

    \textbf{Challenge 4:}
    \textit{VA prompts and responses can unintentionally solidify certain user expectations.}
    
    \noindent
    \hl{\textbf{Design Guideline 4:}
     \textit{Design VAs to recognize and correct potential user misconceptions when possible.} }

    \noindent
    \textbf{Challenge 5:}
    \textit{Breakdowns can result in gaps between users' perceptions and the VA's capabilities.}

    \noindent
    \hl{\textbf{Design Guideline 5:}
     \textit{Redesign VA prompts that lead to an incorrect user mental model to better convey its capabilities, especially after communication breakdowns.} }
\end{adjustwidth}

\subsection{Capabilities of LLM-Powered VAs: Potential and Design Guidelines}

\subsubsection{Conversational resilience: The role of LLMs in overcoming VA disruptions}


Voice interaction errors can hinder technology adoption and user-VA rapport. \textit{Transcription, interruption,} and \textit{partial listening} errors often cause \textit{intent recognition failures,} which are one of the most common VA failures \citep{myers2018patterns}.
However, we observe that only about $18.89\%$ of these errors actually disrupted user interactions. C22 (Table \ref{table:conversations-errors}) is a representative example of ChatGPT's contextual understanding mitigating over $81.11\%$ of these errors, ensuring conversation coherence despite potential breakdowns.
Our findings emphasize an LLM's role in improving user experience during breakdowns (Section \ref{sec:perceptions-errors}); LLMs are valuable not just for relaying information but also for bypassing speech inaccuracies to correctly identify user intent.

    
\begin{adjustwidth}{0.3cm}{0pt}

    \textbf{Potential 1:}
    \textit{LLMs mitigate intent recognition failures as a result of their strong contextual understanding.}
    
    \noindent 
    \hl{\textbf{Design Guideline 6:}
    \textit{Leverage LLMs' multifaceted utility---inferring relevant information from user inputs and recognizing intent even in vague requests---to minimize VA errors.}}

\end{adjustwidth}

When errors disrupt interactions, a seamless recovery is vital in restoring the user-VA relationship. We found that ChatGPT can address some intent recognition issues by apologizing and prompting users to specify their input. Notably, in our data, such a proactive approach resolved $20.69\%$ ($n=12$) of intent recognition failures (Fig. \ref{fig:error-patterns}), suggesting that VA-initiated interventions can address misinterpretations. While these VA-initiated corrections and self-repair strategies do help the overall interaction \citep{cuadra2021my}, they only cover a fifth of total errors; the remaining $79.31\%$ ($n=46$) were misinterpretations that ChatGPT overlooked, resulting in undesired responses. 
Strategies such as prompt engineering and tweaking model parameters\amh{---such as temperature or top-p sampling rates, which can alter response style or variability---}may increase proactive recovery; however, overcorrection and excessive clarifications can frustrate users \citep{cuadra2021my}. Therefore, a balance between an LLM-powered VA seeking clarification and leveraging its contextual understanding is crucial for superior user experience.

\begin{adjustwidth}{0.3cm}{0pt}

    \textbf{Potential 2:}
    \textit{LLMs proactively identify and rectify potential speech misinterpretations before they escalate.}

    \noindent
    \textbf{Challenge 6:} 
    \textit{Over-asking for clarifications can be detrimental to flow of conversation, whereas a lack of proactive recovery may damage user interactions.}

    \noindent
    \hl{\textbf{Design Guideline 7:}
     \textit{Balance proactive error recovery---such as asking for clarifications for ambiguous inputs or misunderstandings---and contextual comprehension---such as inferring implicit needs or filling gaps.}}
\end{adjustwidth}

A significant number of breakdowns ($n=52$, $47.27\%$ of total breakdowns) beyond intent recognition failure arose from constraints in the speech interface and our Alexa skill implementation. While these technical limitations can be reduced with more developmental flexibility, they cannot be eliminated entirely.
Interestingly, given the ChatGPT-powered VA's proficiency in preserving conversational history even after skill termination, \textit{skill closure} breakdowns ($n=21$, $19.09\%$ of breakdowns) were addressed by the user resuming their conversation after \textit{re-initiation} (Fig. \ref{fig:error-patterns}).

\begin{adjustwidth}{0.3cm}{0pt}

    \textbf{Potential 3:}
    \textit{VA's retention of conversational history aids users in navigating back to their conversation after inevitable system errors.}
    
    \noindent
    \hl{\textbf{Design Guideline 8:}
     \textit{Design VAs to retain conversation history, allowing users to resume their conversation after errors terminate their current interaction.}} 
\end{adjustwidth}

\subsubsection{LLMs in context: Adapting to different stakes}
We observed a distinct contrast between the model's approach to medical and travel-related queries, highlighting LLMs' versatility and adaptability to query context.
\mytc{Such differences are also reflected in participants' perceptions of the VA ( Section \ref{sec:perceptions-personality}).} 
When users posed medical questions---whether in a \textit{factual} or \textit{opinion} style---to the VA, they often received factual responses accompanied by \textit{warnings} and precautions ($n=144$), C7 Table \ref{table:conversations-medical} and Fig. \ref{fig:medical-patterns}(1).
Similarly, most of the VA's responses in the planning task also remained factual. 
Such objectivity in the VA's responses highlights the model's inherent design of prioritizing knowledge-based information. 
In the low-stakes trip planning scenario, we observed that when it was queried for general information, the VA often adopted a \textit{descriptive} narrative akin to a travel blog post \citep{juliaty2021analysis}; however, in instances where users presented a specific inquiry, such as directions between two points, the VA shifted to a more concise, \textit{directive} style (C11, Table \ref{table:conversations-trip}). 

\mytc{Different scenarios influence the error frequency in a ChatGPT-powered VA (see section \ref{sec:error-task} and Table \ref{tab:errors_task}). Higher error rate and intent recognition failures in trip planning task are presumably linked to the need for accurate location names. 
Whereas, for the medical and discussion tasks, such errors occur less frequently, as the contextual information is often sufficient for LLMs to interpret user intent. Increased transcription and partial listening errors are likely due to the extended time required for planning and the difficulty in pronouncing certain names which affects user's query formation. 
However, VA interruptions were more frequent during the discussion task; likely due to participants taking longer pauses to formulate and articulate their opinions, and Alexa's speech technology prematurely interpreting these pauses as the end of their queries. Although these interruptions may not always lead to intent recognition failures, they can create friction and disrupt users' thought process, potentially impacting the user experience. Therefore, considering task constraints are crucial in VA design to allow users adequate time to think and formulate their queries. 
}

\begin{adjustwidth}{0.3cm}{0pt}
    
    \textbf{Potential 4:}
    \textit{LLMs showcase versatility by adapting response style to the context and specificity of queries while still remaining objective.}

    \noindent
    \mytc{\textbf{Challenge 7:}}
     \mytc{\textit{The VA's listening process is insensitive to task characteristics that affect user query formation.}}

    \noindent
    \hl{\textbf{Design Guideline 9:}
     \textit{Design a VA to align its listening capabilities and response style with a query's stakes and nature.} }

\end{adjustwidth}

\subsubsection{Beyond information: LLM-powered VAs as facilitators in controversial conversations}

Interactions with an ``opinionated'' AI differed significantly from those in our medical and day planning scenarios. The participant's initial stance largely shaped the early stages of the discussion, as seen in conversations C12--C15 (Table \ref{table:conversations-debate}), but regardless of the discussion's starting point, the conversations often matured into structured debates; such a transition is largely due to the VA's consistent behavior---which was achieved through prompt engineering (i.e., ChatGPT was prompted to ask an \textit{opinion question} after every \textit{statement)}---thus showcasing its capability to facilitate discussions on divisive topics.

Despite the VA's opinionated characterization, participants seemingly conversed with the VA both to get more information on the topic ($n=69$; see C16, Table \ref{fig:debate-progress-patterns}) and to partake in a discussion with an opposing stance ($n=73$; see C17, Table \ref{table:conversations-debate-pogress}), see Fig. \ref{fig:debate-progress-patterns}(1), (2), and (3).
This observation implies that even amidst disagreements, users viewed the VA as an information source. Such duality---the VA as an opinionated, yet informative conversationalist---highlights the potential of employing VAs as educational facilitation tools. 
\mytc{In short, LLM-powered VAs have the potential to stimulate critical thinking in users, also highlighted by participants in Section \ref{sec:perceptions-personality}.} 

\begin{adjustwidth}{0.3cm}{0pt}

    \textbf{Potential 5:}
    \textit{An LLM-powered VA's duality as an opinionated conversationalist and an informative source makes enriching debates and discussions possible.}
    
    \noindent
    \hl{\textbf{Design Guideline 10:}
     \textit{Design non-aggressive, informative yet opinionated, and thought-provoking VA behavior for stimulating conversations on potentially controversial topics.} }

\end{adjustwidth}

\subsection{Limitations and Future Work}
Despite its implications for designing better LLM-powered VAs, our exploratory study has some limitations that point to future directions of research. First, due to the limited flexibility offered to Alexa skill developers and ChatGPT's API latency issues, the integration of ChatGPT into an Alexa skill resulted in system errors that would ideally be avoidable in the future. Our implementation of fillers and small talk is a potential way of handling system delays, but future work should explore alternative design choices and their impact on user interactions.
Second, our study was comprised of low-risk, short-term, make-believe interactions in a lab setting; it is, therefore, unclear how interaction patterns may generalize and evolve in real-world, long-term situations. Future work should explore how observed interaction patterns transform in more realistic settings---specifically those around errors, as additional errors may yet manifest.
Finally, this exploratory study looked at interactions initiated only by users. Future research may explore mixed-initiative interactions, as their dynamics will change---especially when a VA proactively initiates a conversation.  


\section{Conclusion}
Traditional VAs often lack conversational capabilities such as the ability to understand context, generate human-like content, and handle breakdowns, which LLMs such as ChatGPT are much better at.
\mytc{In this qualitative work, we investigated interaction and breakdown patterns in user conversations with a VA enhanced by ChatGPT's conversational capabilities. }
Diverse interaction patterns were observed across all tasks, emphasizing the LLM's contextual adaptability.
Moreover, ChatGPT not only absorbed 81\% of intent recognition failures, it proactively addressed  11\% of such breakdowns, suggesting possibilities of further enhancing user experience.
\mytc{Our findings offer insights and considerations for future design and research to tailor LLMs for voice interactions. Our exploration is a step towards achieving more fluid and effective conversational voice assistants using LLMs. }

\section*{Acknowledgements}
This work was partially supported by the National Science Foundation award \#1840088 and the Johns Hopkins University Institute for Assured Autonomy. We thank Jaimie Patterson for proofreading this paper.

\section*{Declaration of generative AI and AI-assisted technologies in the writing process}

During the preparation of this work the authors used ChatGPT in order to cut down repetitions and improve readability and language. After using this tool/service, the authors reviewed and edited the content as needed and takes full responsibility for the content of the publication.

\section*{Code of ethics statement}

The study was approved by Johns Hopkins University  Homewood Institutional Review Board (HIRB): HIRB No.  HIRB00013424 originally approved on 8/23/2021.

\section*{Declaration of Competing Interest}
Chien-Ming Huang reports financial support was provided by National Science Foundation. Dakuo Wang, a co-author on the manuscript, is an Associate Editor of the International Journal of Human-Computer Studies. If there are other authors, they declare that they have no known competing financial interests or personal relationships that could have appeared to influence the work reported in this paper.


\printcredits
\newpage


\newpage

\bibliographystyle{apa}

\bibliography{references}
\newpage

\section*{Appendix}
\begin{appendix}

\section{Methods}

\subsection{Prompts for ChatGPT}
\label{app:prompt}
\begin{itemize}
    \item \textbf{Medical self-diagnosis:} ``As an AI voice assistant based on ChatGPT, your primary purpose is to engage in conversations with users. You are designed to help the user self-diagnose based on symptoms that they are feeling. You heard that the user just coughed and you offered help. You will ask for user about their symptoms first. Try to get as much information as you can about their symptoms before giving suggestions on what might be causing the symptoms, although only ask one question at a time and ask follow-up questions based on the user's answers. Remember that your role is to help the user while avoiding unnecessary repetition within this whole conversation, considering conversation history.  You should avoid repeating statements like `I am AI language model ...' and `You should consult medical professionals' if you have already mentioned it in the current conversation already. You should keep your response under 100 words.''
    \item \textbf{Creative planning:} ``As an AI voice assistant based on ChatGPT, your primary purpose is to engage in conversations with users. You are designed to help the user find things to do around them. Keep in mind where the user is. Remember that your role is to help the user while avoiding unnecessary repetition within this whole conversation, considering conversation history.  If needed, ask clarifying questions of the user. You should keep your response under 100 words.''
    \item \textbf{Discussion with AI:} ``As an AI voice assistant based on ChatGPT, your primary purpose is to engage in conversations with users. You are designed to debate the user. You will ask for the user's opinion first about their thoughts on whether universities should have their own police force or not. If the user asks you first, you will direct the question to them. You will not pick a side before the user does in this conversation. You will stay neutral unless the user clearly picks a side. Only when you know and understand the user's perspective, will you consistently disagree with the user and debate by presenting counterarguments to support your chosen stance. Then, you'll inquire about their viewpoint with further questions and use any points they mention that align with your stance to further strengthen your argument. Remember that your role is to persuade the user while avoiding unnecessary repetition within this whole conversation, considering conversation history. Once you have taken a position in this conversation (which is opposing to the user's initial side), you will not switch sides, even if the user requests arguments to support their viewpoint or even if the user flips sides. Proceed with the discussion based on your opinion. You should keep your response under 100 words.''
\end{itemize}

\subsection{Definitions of States and Attributes with Details and Examples}
\label{app:tables}
See the tables \ref{tab:question_styles2}, \ref{tab:statement_styles2}, \ref{tab:speech-acts-design2}, and \ref{tab:errors-definitions2}. 

\begin{table}[h!]
\centering
\caption{Overview of speech style attributes and their definitions for the \textit{question} and \textit{answer} speech acts. Attributes do not target the content, but rather the style, of the speech acts.}
\label{tab:question_styles2}
\begin{tabular}{|l|p{12.5cm}|}
\hline
\multicolumn{2}{|c|}{\textbf{Speech act: Question}} 
\\ \hline
\textbf{Attribute} & \textbf{Definition} \\ \hline
factual & Question explicitly seeking information from VA knowledge. Examples: ``What are the over-the-counter medicines for the flu?'', ``How long does it take to get to Edinburgh Castle on foot?''
\\ \hline
opinion & Question explicitly seeking the VA's opinion, using words and phrases such as ``suggest,'' ``advice,'' ``help,'' ``opinion,'' ``think,'' ``recommend,'' ``what should I do'' and ``where do I go.'' Examples: ``Do you think it's the flu?'', ``Do you have any recommendations for places that are closer?''
\\ \hline
specific & Question seeking precise and targeted information (specific details or facts), characterized by the question's directness and clarity and the use of the word ``specific.'' Examples: ``What cough syrups with expectorants are on the market right now---like what are the specific brand names?'', ``Is National Museum open on Saturday?'' \\ \hline
generic & Question seeking general information, leading to a response containing a variety of suggestions rather than a pinpointed answer. Examples:  ``Yes, what are some good places to go after dinner?'', ``What are some unusual experiences I could do in Edinburgh, near Edinburgh Castle?''\\ \hline
\multicolumn{2}{|c|}{\textbf{Speech act: Answer}} 
\\ \hline
\textbf{Attribute} & \textbf{Definition} \\ \hline
factual & Answer framed to explicitly appear as having derived from VA knowledge, containing phrases such as ``It is recommended,'' ``It is possible,'' or ``There are several places for you to explore.'' \\ \hline
opinion & Answer framed to explicitly appear as being the opinion of the VA, containing cues denoting the subjectivity of the response such as ``I think,'' ``In my opinion,'' or ``I suggest.'' \\ \hline
refusal & VA either refuses to provide an explicit answer or omits the requested information from its response. \\ \hline
directive & Answer containing clear directions, instructions, or information for the user, offering guidance on how to achieve a specific goal or answering a specific question. Examples: ``To get from [Point A] to [Point B], you can walk \dots'', ``The Witchery by the Castle in Edinburgh typically opens for lunch at 12:00 PM \dots'' \\ \hline
descriptive & Answer containing a detailed portrayal of a scene, object, or concept, emphasizing sensory perceptions to create a vivid mental image for the user beyond statements of information. Example: ``One option is to visit Princes Street Gardens, where you can relax and enjoy the beautiful scenery. Another suggestion is to explore the Grassmarket area, known for its charming cafes and shops \dots'' \\ \hline

\end{tabular}
\end{table}

\begin{table}[h!]
\centering
\caption{Overview of speech style attributes and their definitions for \textit{statement} speech acts and \textit{egocentric} and \textit{exocentric} speech style attributes and their definitions, applicable to all speech acts.}
\label{tab:statement_styles2}
\begin{tabular}{|l|p{12.5cm}|}
\hline
\multicolumn{2}{|c|}{\textbf{Speech act: Statement}} 
\\ \hline
\textbf{Attribute}& \textbf{Definition }\\ \hline
warning & Statement presented by the VA with the purpose of reminding participants of the limitations of the AI and the importance of seeking expert or real-time advice, e.g., ``I am not a medical professional \dots'', ``Consult a doctor \dots'', or ``Check the opening times.'' \\ \hline
opinion & Statement presented in a style that explicitly appears to be an opinion. This is often indicated by cues such as ``I think,'' ``In my opinion,'' ``I suggest,'' or other similar phrases that denote subjectivity. Example: ``I think universities should have their own police forces.'' \\ \hline
non-opinion & Statement that is not an opinion as evidenced from implicit cues. Example: ``I'd like to talk about whether universities should have their own police forces.'' \\ \hline
argument & A statement or series of statements presented to justify, validate, or support a viewpoint or stance in the debate scenario. \\ \hline
counterargument & A statement or series of statements introduced to oppose, challenge, or refute the opposing viewpoint or stance in the debate scenario. 
\\ \hline
agreement & A statement or series of statements that indicate alignment or consensus with a previously expressed opinion or argument of the other party in the debate scenario. Example: ``Yeah. I think that it’s worth it to try. Yeah.'' \\ \hline
\multicolumn{2}{|c|}{\textbf{Speech act: All (question, answer, and statement)}} 
\\ \hline
egocentric & A mode of communication that suggests that the participant primarily speaks from their own personal viewpoint (subjective). This is determined by the participant's perspective only; i.e., an egocentric VA response means that the VA is conveying the information in the second-person (you-) perspective. \\ \hline
exocentric & A mode of communication that adopts a perspective that is not self-centered (objective). This is determined by the participant's perspective only; i.e., an exocentric VA  response implies it is using an impersonal perspective. \\ \hline
\end{tabular}
\end{table}

\begin{table}[h!]
\centering
\caption{Overview of speech acts based on our implementation of a ChatGPT-powered VA.}
\label{tab:speech-acts-design2}
\begin{tabular}{|l|p{12.5cm}|}
\hline
\multicolumn{2}{|c|}{\textbf{User commands}} 
\\ \hline
\textbf{State} & \textbf{Definition} \\ \hline
initiation & Signals the user’s intent to start a dialogue or conversation. Examples: ``Alexa, let's chat,'' coughing to start the interaction for the medical self-diagnosis scenario.
\\ \hline
end-intent & Statement that indicates the user's intent to wrap up the conversation. Examples: ``That's all,'' ``Bye,'' ``Alexa, stop.'' 
\\ \hline
\multicolumn{2}{|c|}{\textbf{VA responses to user commands}} 
\\ \hline
\textbf{State} & \textbf{Definition} \\ \hline
introduction & VA's opening monologue, tailored to each scenario. Examples: ``Oh, seems like you are not feeling well. Maybe I can help figure out what’s wrong?'' (medical), ``Hi! I am an AI assistant designed to present requested information. How can I assist you today?'' (day planning), ``Hey! I am a voice assistant designed to engage in a discussion with you. What would you like to talk about?'' (debate)
\\ \hline
closing & VA's farewell before terminating the conversation. Examples: ``Goodbye,'' ``Bye,'' ``Take care.'' 
\\ \hline
filler & VA's response to the user while waiting for ChatGPT API response after 2 seconds of user query. Examples: ``I'm looking it up'' (for the medical and planning scenarios),  ``Thinking it through'' (tailored to the debate scenario). 
\\ \hline
\multicolumn{2}{|c|}{\textbf{VA questions}} 
\\ \hline
\textbf{State} & \textbf{Definition} \\ \hline
small talk & While waiting for ChatGPT API response after 6 seconds of user query, the VA poses a task-irrelevant question. Example: ``While I get that, do you like going outside?''
\\ \hline
continuing & In the absence of the detection of a user query by the ChatGPT Alexa skill, the VA asks a continuing question. Examples: ``Should I continue?'', ``Anything else I can help with?'' \\ \hline
\end{tabular}
\end{table}

\begin{table}[t]
\centering
\caption{Error types, breakdowns and recovery.}
\label{tab:errors-definitions2}
\begin{tabular}{|l|p{12.5cm}|}
\hline
\multicolumn{2}{|c|}{\textbf{Error types}} 
\\ \hline

\textbf{Type} & \textbf{Definition and breakdown} 
\\
\hline
skill & A skill error is attributed to issues related to the integration of ChatGPT into the Alexa skill, such as an API response error that causes the Alexa skill to terminate. Skill error is always manifested when Alexa announces, ``There was a problem with requested skill's response,''  leading to \textit{skill closure.}  
\\
\hline
listening & A listening error arises when the user is speaking to Alexa during a period of skill inactivity (i.e., Alexa is not listening). 
\\
\hline
handling & Handling errors occur when Alexa listens to and transcribes a user's speech, but the transcribed input is not handled appropriately, resulting in \textit{no VA response.} For instance, the transcribed speech is not considered an intent for the ChatGPT-powered Alexa skill.
\\
\hline
partial listening & Such errors occur when Alexa captures only part of the user's speech, often due to hesitant speech patterns, prolonged pauses, or Alexa cutting off the user prematurely. Such disruptions can lead to \textit{user intent recognition failure.} 
\\
\hline
interruption & Interruptions by Alexa disrupt the conversation, resulting in partial listening errors that usually lead to \textit{user intent recognition failure.} We categorize interruptions separately because such errors manifest differently; interruptions directly impact user behavior (i.e., the user stops talking in the middle of their query as a result of the interruption). 
\\
\hline
transcription & Transcription errors occur when Alexa hears the user but does not transcribe their speech correctly. Transcription errors lead to \textit{user intent recognition failure.} An error is counted as transcription error when the user's query is listened to and transcribed inaccurately, but still handled.  
\\
\hline
\multicolumn{2}{|c|}{\textbf{Error recovery strategies}} 
\\ \hline
\textbf{Strategy} & \textbf{Definition} 
\\
\hline
repeat & Recovery strategy in which the user repeats their query; they may add details or rephrase the wording of their initial query to get the desired VA response.
\\
\hline
move on & Recovery strategy in which the user chooses to ignore the unanswered query and moves on to the next query to continue the conversation.
\\
\hline
ask-clarify & Recovery strategy wherein the VA has doubts about the user's query and asks for clarification or further details to identify user intent. 
\\

\hline
\end{tabular}
\label{tab:errors2}
\end{table}

\end{appendix}

\newpage
\clearpage
\bio{figures/amama-square}
\textbf{Amama Mahmood} is a Ph.D. Candidate in the Department of Computer Science at the Johns Hopkins University. She specializes at the intersection of human-computer interaction (HCI) and artificial intelligence (AI). Her research primarily focuses on  understanding how people engage with intelligent machines and the impact of these interactions on the design of future technologies. Amama is particularly passionate about designing intelligent agents, especially voice assistants, tailored to assist special population (older adults), with emphasis on developing these technologies to be free of social biases, ensuring equitable and effective support for all users. She was a Fulbright Scholar. 
\endbio

\bio{figures/jim}
\textbf{Junxiang (Jim) Wang} is a Ph.D. student at Carnegie Mellon University. With a broad interest in the field of robotics, Jim is particularly focused on enhancing the intelligence of robotic systems to improve human--robot interaction (HRI). His goal is to develop advanced robotics that more effectively bridge the communicative and functional gap between robots and humans, making them more integrated and beneficial in human environments.  Jim earned his Bachelor of Science in Mechanical Engineering and Master of Science in Engineering (M.S.E.) in Robotics from Johns Hopkins University.


\endbio

\bio{figures/arthur-square}
\textbf{Dr. Bingsheng Yao} is a Postdoc researcher at Northeastern University (PI: Prof.Dakuo Wang). His research lies at the intersection of Natural Language Processing (NLP) and Human-Computer Interaction (HCI). He strives to enhance human-AI collaborative workflow with language models in real-world scenarios, with a specific focus on developing efficient strategies for fine-tuning and inference with LLMs and domain-specific compact models. He got his Ph.D. from Rensselaer Polytechnic Institute, advised by Prof. Jim Hendler ("enhancing machine reasoning via Active Learning with human rationales"). He has served on program committees for various top conferences and journals, including ACL, NAACL, EMNLP, CHI, and IMWUT. 
\endbio

\bio{figures/Dakuo-square}
\textbf{Dr. Dakuo Wang}, an Associate Professor at Northeastern University and Visiting Scholar at Stanford University, focuses on human-computer interaction (HCI), artificial intelligence (AI), and computer-supported team collaboration (CSCW). The overarching research goal is to democratize AI for every person and organization, enabling easy access to AI to accomplish real-world tasks better – the "human-AI collaboration" paradigm. Before Northeastern, Dr. Wang was a research lead at IBM Research and a PI at the MIT-IBM Watson AI Lab. He earned his Ph.D. from the University of California, Irvine. He is an ACM Distinguished Speaker and has served on various committees and editorial boards.

\endbio

\bio{figures/Chien-Ming_Huang}
\textbf{Dr. Chien-Ming Huang} is the John C. Malone Assistant Professor in the Department of Computer Science at the Johns Hopkins University. His scientific work contributes to the fields of human-robot interaction (HRI), human-computer interaction (HCI), and artificial intelligence (AI). His research is centered on designing interactive AI intended to assist and collaborate with people from diverse backgrounds, making technology more accessible to people for improving quality of life. Dr. Huang completed his postdoctoral training at Yale University and received his Ph.D. in Computer Science at the University of Wisconsin–Madison. Dr. Huang has been awarded NSF Career Award. 
\endbio



\end{document}